\documentclass[11pt,a4paper]{amsart}
\usepackage{geometry}                
\usepackage{graphicx}
\usepackage{amssymb}
\usepackage{epstopdf}
\usepackage{fullpage}
\usepackage{color}
\usepackage{float}
\usepackage{caption}
\usepackage[table]{xcolor}
\usepackage{tablefootnote}

\usepackage{url}

\definecolor{mattgreen}{RGB}{216,228,188}
\definecolor{mattred}{RGB}{218,150,148}
\definecolor{mattgrey}{RGB}{64,64,64}

\def\mybar#1{
  {\color{mattgrey!50}\rule{#1in}{8pt}}}

\usepackage{graphicx} 

\usepackage{array} 
\usepackage{paralist} 

\begin{document}

\title{Measures of Human Mobility Using Mobile Phone Records Enhanced with GIS Data}

\author{Nathalie E. Williams}
\address{Department of Sociology and Jackson School of International Studies, University of Washington, Box 353340,Seattle, WA 98195}
\email{natw@uw.edu}

\author{Timothy A. Thomas}
\address{Department of Sociology, University of Washington, Box 353340,Seattle, WA 98195}
\email{t77@uw.edu}

\author{Matthew Dunbar}
\address{Center for Studies in Demography and Ecology, University of Washington, Box 353412, Seattle, WA, 98195}
\email{mddunbar@uw.edu}

\author{Nathan Eagle}
\address{Department of Epidemiology, Harvard University, Boston, MA 02115}
\email{nathan@mit.edu}

\author{Adrian Dobra}
\address{Department of Statistics, Department of Biobehavioral Nursing and Health Systems, Center for Statistics and the Social Sciences and Center for Studies in Demography and Ecology, University of Washington, Box 354322, Seattle, WA 98195}
\email{adobra@uw.edu}

\date{\today}                                           

\begin{abstract}
In the past decade, large scale mobile phone data have become available for the study of human movement patterns. These data hold an immense promise for understanding human behavior on a vast scale, and with a precision and accuracy never before possible with censuses, surveys or other existing data collection techniques. There is already a significant body of literature that has made key inroads into understanding human mobility using this exciting new data source, and there have been several different measures of mobility used. However, existing mobile phone based mobility measures are inconsistent, inaccurate, and confounded with social characteristics of local context. New measures would best be developed immediately as they will influence future studies of mobility using mobile phone data. In this article, we do exactly this. We discuss problems with existing mobile phone based measures of mobility and describe new methods for measuring mobility that address these concerns. Our measures of mobility, which incorporate both mobile phone records and detailed GIS data, are designed to address the spatial nature of human mobility, to remain independent of social characteristics of context, and to be comparable across geographic regions and time.  We also contribute a discussion of the variety of uses for these new measures in developing a better understanding of how human mobility influences micro-level human behaviors and well-being, and macro-level social organization and change.
\end{abstract}

\maketitle

\tableofcontents

\section{Introduction}

Human mobility, or movement over short or long spaces for short or long periods of time, is an important yet under-studied phenomenon in the social and demographic sciences.  While there have been consistent advances in understanding migration (more permanent movement patterns) and its impact on human well-being, macro-social, political, and economic organization \cite{Donato-1993,Durand-et-al-1996,Harris-Todaro-1970,Massey-1990,Massey-et-al-1993,Massey-Espinosa-1997,Massey-et-al-2010,Stark-Bloom-1985,Stark-Taylor-1991,Taylor-1987,Todaro-1969,Todaro-Maruszko-1987,VanWey-2005,Williams-2009}, advances in studies of mobility have been stymied by difficulty in recording and measuring how humans move on a minute and detailed scale.  This gap is particularly glaring given that mobility is likely a fundamental factor in behavior and macro-level social change, with likely associations with key issues that face human societies today, including spread of infectious diseases, responses to armed conflict and natural disasters, health behaviors and outcomes, economic, social, and political well-being, and migration.  In this context, new methods for measuring human mobility could lead to significant advances in the policy relevant social and demographic sciences.

Mobile phone data have recently become available for the study of human mobility.  Such data are continuously collected by wireless-service providers for billing purposes and to improve the operation of their cellular networks \cite{becker-et-al-2013}. Every time a person makes a voice call, sends a text message or goes online from their mobile phone, a call detail record (CDR) is generated which records time and day, duration and type of communication, and an identifier of the cellular tower that handled the request. The approximate spatiotemporal trajectory of a mobile phone and its user can be reconstructed by linking the CDRs associated with that phone with the locations (latitude and longitude) of the cellular towers that handled the calls. This exciting new type of data holds immense promise for studying human behavior with precision and accuracy on a vast scale never before possible with surveys or other data collection techniques \cite{tatem-2014}.  As mobile phone penetration increases dramatically worldwide to an estimated 120.8 (90.2) mobile-cellular subscriptions per 100 inhabitants in developed (developing) countries by the end of 2014 \cite{ITU-2014}, selection in who uses mobile phones is decreasing, thereby reducing biases related to phone ownership \cite{Weslowski-et-al-2013b} and making CDRs ever more appropriate for studying human mobility of whole populations.

There is a significant body of literature that has already made key inroads into understanding mobility using this exciting new data source, and there have been several different measures of mobility used \cite{Blumenstock-2012,Blumenstock-Eagle-2012,Gonzales-et-al-2008,Lu-et-al-2012,Phithakkitnukoon-et-al-2012,Song-et-al-2010,Weslowski-et-al-2012,Weslowski-et-al-2013a,Weslowski-et-al-2013b}.  However, there has been little discussion and assessment of these measures. Consequently, we understand little about what they actually measure and how they perform.  Indeed, we argue that existing measures of mobility from CDRs do not measure mobility accurately or consistently, are confounded with other contextual characteristics, and are therefore not suitable to advance mobility studies.  We further argue that the need for improved measures of mobility would be best addressed immediately as this will influence the conclusions of future studies of mobility using mobile phone data.

Towards developing accurate and meaningful measures of mobility with CDRs, and advancing this promising area of social science, in this article we propose six novel measures of mobility derived from CDRs.  We define key dimensions of mobility and describe existing measures of mobility and the problems they entail.  Using this background, we then propose and analyze six new measures that directly address each dimension of mobility and overcome the inherent problems with existing measures by combining CDRs with detailed GIS data on road networks.  We carefully assess our measures using CDR and GIS data from Rwanda.  An important difference in our proposed measures from those used previously is that they are fundamentally based on existing spatial analytical methods, reflecting the spatial nature of mobility.  A second key difference is that they account for how humans actually move, which is most often via road networks and through many places, instead of by apparition or ``as the crow flies'' from one place to another.  A consequence of our spatial and movement perspectives is that they produce pure measures that address only movement of humans and are not affected by other characteristics of social context besides the roads upon which people move. Another consequence is that they are designed to be broadly applicable to different geographic settings regardless of human behavioral patterns or variation in context.  This article ends with a discussion of the new ways in which these measures can be used to advance the scientific study of human mobility.

For illustration we analyze anonymized CDRs provided by a major cellular phone service provider in Rwanda. These data comprise all mobile phone activity in the provider's network between June 1, 2005 and January 31, 2009 \cite{Blumenstock-2012,Blumenstock-Eagle-2012}. To evaluate existing and new measures of mobility, we define spatiotemporal trajectories of each caller in the provider's network from the CDRs they generate in every given month. This yields 20,139,971 person months of spatiotemporal trajectories \---- for additional details, see {\bf SI Appendix, Section SI2}. Calculation of our measures from these spatiotemporal trajectories is detailed in {\bf SI Appendix, Section SI3.2} which provides formulas and technical details.

\section{Dimensions of Mobility}

In order to better define the problems with existing measures of mobility, to design new measures, and to assess measures, we delineate two key dimensions of mobility.  The first key dimension is the frequency of movement, and  represents the number of times a person goes anywhere.  The higher the frequency, or more times a person moves, the higher should be the value of their mobility measure. What constitutes going somewhere and what designates separate trips depends on definition and these definitions vary by study and context \cite{calabrese-et-al-2013,chen-et-al-2014}. One of our primary motivations is to create coherent measures of mobility, including frequency of movement, that are meaningful in each specific context but comparable across contexts. The second key dimension of mobility is spatial range, or how far a person moves. The further a person moves, the higher should be the value of their mobility measure. 

\section{Existing CDR-Based Measures of Mobility}

Existing measures of mobility derived from CDRs include number of towers used (NTU), distance traveled-straight line (DT-SL), maximum distance traveled (MDT), and the most commonly used measure, radius of gyration (RoG) \--- see, among others \cite{Weslowski-et-al-2013b,Blumenstock-2012,Blumenstock-Eagle-2012,Gonzales-et-al-2008,Lu-et-al-2012,Phithakkitnukoon-et-al-2012,Song-et-al-2010,Weslowski-et-al-2012,Weslowski-et-al-2013a}. Measures of mobility are defined with respect to a fixed period of time, e.g. hours, days, weeks, months or years. Here we chose months as the reference time period, but our methodological developments and conclusions are relevant for shorter or longer reference time periods. The NTU measure counts the number of cellular towers from which a person called in the requisite period of time.  The DT-SL measure, also called average travel distance \cite{Lu-et-al-2012}, is the sum of straight line or ``as the crow flies'' distances between towers from which consecutive calls or texts were made. The MDT measure calculates the maximum straight line distance between two towers that a person used.  The RoG is determined by first finding the center of mass of all cellular towers that a person used. The straight line distances from the center of mass to each used tower are calculated, and the value of RoG is the square root of the mean of the squares of these distances. {\bf SI Appendix, Section SI3.1} gives formulas and related details.

\begin{figure}[htbp!]
\begin{center}
\centerline{\includegraphics[width=.7\textwidth]{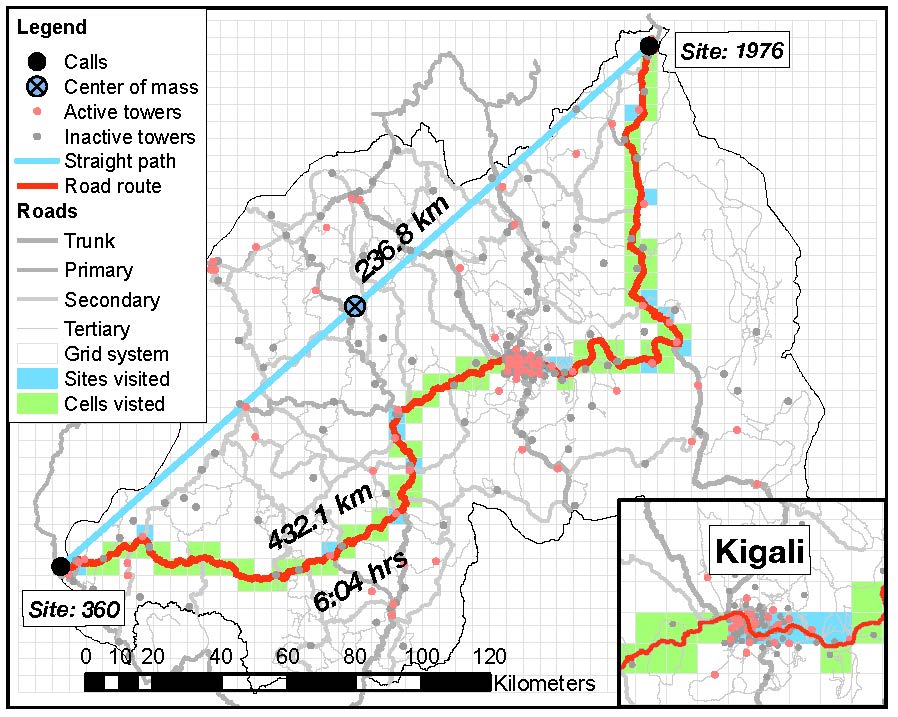}}
\caption{Map of the monthly spatiotemporal trajectory of the caller with the largest monthly RoG. This caller which we refer to as $\mathcal{P}$ made two calls in October 2005: the first one from a cellular tower located in the grid cell labeled ``Site 1976" and the second one from a cellular tower located in the grid cell labeled ``Site 360." There are 2040 5km x 5km grid cells indexed from 1 (the cell in the lower left corner) to 2040 (the cell in the upper right corner). A site is a grid cell that contains at least one cellular tower. The map shows the location of the straight path between sites 1976 and 360, and also the location of the quickest road route \--- the road route with the smallest estimated travel time \--- between the two sites. The straight path between the two towers used by $\mathcal{P}$ is 236.8 km long, while the straight line path between the centroids of sites 1976 and 360 is 237.2 km long. The quickest road route between the centroids of sites 1976 and 360 is 432.1 km long. The estimated travel time along this route is 6 hours and 4 minutes. The map also shows the center of mass required for the calculation of RoG which is located in the middle of the straight path, as well as Rwanda's borders, Rwanda's road network structure with trunk, primary, secondary and tertiary roads, and the locations of the all the 239 cellular towers references in the Rwandan CDRs. We note that only 78 of these towers were active (i.e., handled at least one communication) in October 2005. The grid cells that contain at least one active tower in October 2005 are referred to as sites for that month. The visited sites associated with the spatiotemporal trajectory of $\mathcal{P}$ are the sites which are intersected by the quickest road route between sites 1976 and 360. There are 19 visited sites which are shown in blue. All the grid cells intersected by this route are called visited cells. The visited cells that are not visited sites are shown in green. The inset shows the capital Kigali and its surrounding area. This is the region with the highest cellular tower density in Rwanda. The Rwandan road network is publicly available data under the Open Database License, and comes from OpenStreetMap (openstreetmap.org), a global open-source mapping project.\label{fig:personRoG}}
\end{center}
\end{figure}

We exemplify the evaluation of these four measures with the spatiotemporal trajectory of the caller, $\mathcal{P}$, who had the largest RoG from all 20 million trajectories in the Rwandan data \---- see Figure \ref{fig:personRoG}. During October 2005, $\mathcal{P}$ made only two calls in this provider's network: the first call from a location near the northern border with Uganda and the second call from a location near the western border with Democratic Republic of Congo. The NTU measure for this person is equal to 2 (10th percentile). The DT-SL and the MDT measures are both 236.8 km (78th and 100th percentile, respectively). The RoG of $\mathcal{P}$ is 118.6 km.

These four measures of mobility have several critical shortcomings. Two of the primary problems are caused by their direct definition with respect to the location of the cellular towers and the fact that tower placement is not random or evenly spaced \cite{becker-et-al-2013}. Figures \ref{fig:personRoG} and S1 ({\bf SI Appendix}) show the uneven variation in tower placement in Rwanda: the capital Kigali has a high tower density with respect to the rest of the country which comprises mostly rural areas. Consider a person who lives in Kigali with 50 towers within a 5 km radius.  This individual could regularly move within only this 5 km disk, but their CDRs would document them as using 50 towers and their mobility could be then calculated as high.  Compare this person to a second person living in a rural area with only one tower in a 5 km radius of their home.  Even if they move about this 5 km disk as often as our urban individual, the rural individual would only ever use this one tower and thus be classified as not moving anywhere and attain the lowest mobility rating.  Thus, if not taken into account, variations in tower density create variations in mobility that do not actually exist. This issue is further exacerbated by the fact that cellular towers are placed more often in urban areas with high population density, politically important areas, such as capital cities, or wealthy areas with higher mobile phone penetration. In short, because tower density is confounded with  social, economic, political, and demographic characteristics of context, existing mobility measures that rely on tower density are confounded with these contextual factors as well.  

A second, and related, concern is that the placement of cellular towers varies with time.  In many countries, where the mobile infrastructure has not yet reached saturation, new cellular towers are built every year to accommodate increasing numbers of users. For example, the total 269 towers in Rwanda existed in various time periods from June 2005 to January 2009 and Figure S3 ({\bf SI Appendix}) documents the growth in the number of Rwandan callers from 190 thousands in June 2005 to more than 1 million in December 2008. Towers are added in the proximity of other towers, but also in regions without previous cellular coverage \---- see Figures S4 and S5 ({\bf SI Appendix}), while others are taken off the grid. This creates a situation where the spatial density of cellular towers, which is already a problem for existing CDR-based mobility measures, is time-varying.  In other words, there is temporal variance in the spatial variance of cellular towers. Because existing measures use towers as their spatial reference points, this causes a situation of spatial and temporal bias in these measures.

In addition to the problem that existing measures are confounded with tower density, they are also inherently confounded with call frequency.  The more often a person calls, the more towers at which they will be registered.  A person who uses their phone frequently will likely have a different mobility rating, compared to a person with the same spatiotemporal trajectory but lower calling frequency.  This problem is particularly acute given that call frequency is selective of men and wealthier people \cite{Weslowski-et-al-2013b}. Confounding with call frequency is essentially a missing data problem and creates inconsistencies between individuals. An analogous missing data problem is with areas that have no cellular towers. CDRs do not account for movement of people in areas with no tower coverage. Thus this zero tower issue is also a missing data problem, but creates inconsistencies between areas. 

The temporal and spatial sparsity of CDRs \cite{becker-et-al-2013} that affects the call frequency and zero towers problems becomes apparent in the spatiotemporal trajectory of the highest RoG caller $\mathcal{P}$. This person made only two calls near two distant Rwandan borders. Given the time elapsed between the two calls and Rwanda's transportation infrastructure, it is very unlikely that $\mathcal{P}$ traveled by air between the two locations. Therefore $\mathcal{P}$ was most likely present in several other locations in Rwanda, somewhere along the way between the two towers that handled their two calls. This leads to underestimates of the values of NTU, DT-SL and MDT. Even more serious is the fact that locations with cellular coverage that were visited but not represented in the CDRs can have a significant effect in the determination of the location of the center of mass, which subsequently translates into biased values of RoG. 

A fourth problem is that the existing measures of mobility are fundamentally based on implicit, yet unrealistic assumptions about the nature of human movement.  Their definitions involve measurement of distances in straight lines between cellular towers. In fact, humans almost never travel in straight lines and outside of air travel (which we discuss in {\bf SI Appendix, Section SI4}) they do not usually appear in one place, then disappear and appear again in another distant place. For example, with caller $\mathcal{P}$, who was registered as being present in one side of Rwanda, then again in another side of the country, it is likely that he/she traveled longer distances on roads between these points. The straight line distance between the towers used is 236.8 km, but the quickest road route between the same locations measures 432.1 km \--- an increase of 82\%. This clearly causes underestimates in the values of DT-SL and MDT and varying bias in RoG.

The fifth problem is that it is not entirely clear which aspect(s) of mobility most of these measures capture. Due to varying density of cellular towers, the NTU measure does not capture spatial range. However, because it counts unique towers, it also does not assess frequency of movement. The DT-SL measure captures both frequency of movement and spatial range.  The MDT measure, because it incorporates only two of the towers used, captures neither frequency nor spatial range well. The RoG measure does not capture frequency of movement.  While it initially appears to capture spatial range, it does so in an inconsistent manner that is influenced by call frequency from each tower used. Take the example of caller $\mathcal{P}$ (see Figure \ref{fig:personRoG}) and of another fictive caller, $\mathcal{P}^{\prime}$, who makes $1000$ calls from the tower used by $\mathcal{P}$ for their first call, and only one call from the tower used by $\mathcal{P}$ for their second call. The center of mass of $\mathcal{P}^{\prime}$'s trajectory will be very close (236.5 meters away) to the location of the tower used by $\mathcal{P}$ for their first call, and will be $118.1$ km away from the center of mass of $\mathcal{P}$'s trajectory. Thus, despite the spatial range covered by $\mathcal{P}$ and $\mathcal{P}^{\prime}$ being exactly the same, the value of RoG for $\mathcal{P}^{\prime}$ will be 7.5 km which is very small compared to 118.6 km, the value of RoG of $\mathcal{P}$. In summary, three of the four existing measures (NTU, MDT and RoG) do not clearly and consistently measure either of the key dimensions of mobility. Only DT-SL, which incorporates both frequency and spatial range does so.  Yet even this measure suffers from the major shortcomings outlined above.

\section{New CDR-Based Measures of Mobility}

Given these concerns about existing measures of mobility, our intent is to design new measures that: (i) are independent of mobile tower density and the social characteristics of context that influence tower density; (ii) are less dependent on users' call frequency, movement in areas with no tower coverage, and the temporal dynamics of the underlying cellular network of towers; (iii) measure clearly defined aspects of mobility such as the frequency and spatial range of movement; and (iv) are relevant and comparable across contexts, countries, and time.

The first foundation of our measures is a system of grid cells of even size placed across a country or area of study. For Rwanda we chose to work with 2040 grid cells each measuring 5 km x 5 km. Several key explanations related to the practical implementation of a grid system for mobility measurement, including how the grid is placed on a map and the size of grid cells, are discussed in {\bf SI Appendix, Section SI1}.  As shown in Figures \ref{fig:personRoG} and S2 ({\bf SI Appendix}), some grid cells have a cellular tower in them, some do not, and some have multiple cellular towers. We refer to a grid cell with at least one active tower as a site. With the grid system, if an example person, $\mathcal{R}$, calls from a cellular tower, we register them as being located at the centroid of the corresponding site (grid cell).  Movement is then calculated only when $\mathcal{R}$ moves from one site to another.  If $\mathcal{R}$ calls again from another tower in the same site, then they are registered in the same site, and thus have not moved. But if the next call $\mathcal{R}$ makes is handled by a tower in a different site, then they have moved.  Our methodology entirely disposes of cellular towers and instead replaces them with the sites they belong to. By doing so, the problem of spatial variation in tower density is eliminated because grid cells are of even size and non-overlapping. 

By replacing cellular towers with sites, the adverse effect of the temporal variability of the spatial extent of cellular towers coverage is also significantly diminished. In a given time period, a tower is active if it handled at least one cellular communication during that period. Otherwise a tower is inactive and does not contribute to the creation of a site. Figure S5 ({\bf SI Appendix}) shows that, in the Rwandan data, the month to month increase in the number of sites is much smaller than the month to month increase in the number of active towers. Spatiotemporal trajectories constructed with respect to sites instead of cellular towers will have less temporally induced bias as the set of sites will always change less than or equal to the set of active cellular towers during any time period.

The second foundation of our measures is a set of realistic assumptions about how humans travel: they most often use roads, will take the quickest, most accessible road route from one place to another, and the speed of travel is affected by speed limits and quality of road surfaces.  With these assumptions, we use publicly available GIS data on road systems to create routes of travel from one place to another that are not straight lines \--- see {\bf SI Appendix, Section SI1}.  With this information, it is possible to calculate an assumed route of travel between any two points in a country, where the assumed route has the shortest possible travel time compared to all other routes.  Because all our measures are based on a grid system, we create assumed routes of travel that begin at the centroid of a site from which a person placed a call, take the shortest distance route to the nearest road from the site's centroid, travel the quickest route of travel to the site in which their next call was placed, and end at the centroid of that site.

The third foundation of our measures is that humans most often travel on the ground.  Even if they do not make calls at every place they visit, we can assume they existed for some amount of time in every place along a road route, between two subsequent calls.  This assumption partially ameliorates the confounding influence of call frequency and no available towers on mobility measurement.  In the existing measures of mobility, only places where a person made calls are included in the spatiotemporal trajectories these measures are based on, thus higher call frequency inflates mobility ratings.  Here, because we account for places where people made calls and places where they did not but likely existed for any amount of time, call frequency is less confounding.  For spatiotemporal trajectories that involve longer trips with one call at their origin, another call at their destination and no calls in-between (see the example of caller $\mathcal{P}$), the absence of in-between calls has a reduced effect on our proposed measures of mobility because we also include in the trajectory sites and grid cells located on the quickest road routes \--- see Figure \ref{fig:personRoG}. 
	
Based on these foundations we create six new mobility measures, and divide them into three groups depending on which of the two key dimensions of mobility they capture. Group A includes measures that capture the frequency of mobility, but do not capture spatial range; group B includes measures that capture spatial range, but not frequency; and group C includes measures that capture both frequency and spatial range. There is more than one measure in groups B and C and these differ primarily by unit of measurement. The measures within groups are of course related and thus correlate strongly.  Below we describe these new measures, their benefits, and their limitations. We exemplify the evaluation of our new measures with the spatiotemporal trajectory of caller $\mathcal{P}$ who made two calls in October 2005, one call from site 1976 and a second call from site 360 (Figure \ref{fig:personRoG}).

Combinations of measures that belong to every one of these three groups are needed to identify various mobility patterns that exist in a population. For example, caller $\mathcal{P}$ (Figure \ref{fig:personRoG}) made only one long trip, therefore their mobility will be rated as high by measures from group B, but not by measures in groups A and C. Consider two other callers $\mathcal{P}_1$ and $\mathcal{P}_2$ that go from home to work and back for 20 days each month, but $\mathcal{P}_1$'s work is 1 km from his home and $\mathcal{P}_2$'s work is 10 km from her home. Thus they move with equal frequency in a given period of time, but distances between consecutive places in $\mathcal{P}_1$'s trajectory are shorter than those for $\mathcal{P}_2$'s trajectory. In this case, the mobility of $\mathcal{P}_1$ and $\mathcal{P}_2$ will be equal when evaluated by measures from group A, but will differ when evaluated by measures from groups B and C. The particular measure or combination of measures one uses will depend on the research question and context of each study. We advocate at least testing analyses with all six mobility measures.


\subsection{Group A: Measures of frequency of mobility}

{\bf Number of trips (NT)}.  This measure is a count of the number of times a person makes a call from a different grid cell than the previous call. If a person makes a call from one grid cell and their next call is from the same grid cell (regardless if it is from a different tower) then this is not a trip. $\mathcal{P}$ made two calls from two different sites, thus the value of NT is equal to 1 (10th percentile). If $\mathcal{P}$ would have made any number of subsequent calls using only the two towers from site 360, the value of NT would be unchanged.  Note that this measure does not depend on how far a person travels (spatial range): $\mathcal{P}$ could have called from any two of the active sites, and the value of NT would be the same. But if $\mathcal{P}$  makes a call from another site, the value of NT will increase by 1. The amount of time between the calls is disregarded when calculating NT.
	
The limitations of NT come from the incomplete information on mobility contained in CDRs. This measure relies on a specific definition of a trip as a movement between two places where a person existed for any amount of time.  The transportation literature  often defines a trip as movement between two places where a person stayed for a minimum amount of time (often 5 or 10 minutes) \--- see, for example, \cite{chen-et-al-2014}.  Using CDRs, it is not possible to determine how long a person stayed at each place they made a phone call.  This CDR-derived measure could record fewer trips if a minimum time at a destination were required or if a person does not make a call when at a particular destination before leaving for their next destination. More trips would be recorded in cases where a person makes several phone calls when traveling between an origin and destination (or makes longer calls using multiple towers), and no minimum time at a destination were required. However, this limitation is precisely what makes NT comparable across time and place. Definitions of a trip that use any more information than we do here, will necessarily be time and context specific; an intricate definition of what constitutes a meaningful trip in rural Mongolia will certainly be different from what constitutes a trip in New York City. Thus, the limited information that NT uses is both a detraction and a benefit.

\subsection{Group B: Measures of spatial range of mobility} 

The next two measures represent the number of places that a person visited.  Just as with trips, a careful definition of what constitutes a place is required for consistency and comparability across geographic contexts and time.  Both group B measures require an assumption that all places in which a person exists for any amount of time could be important.  Some of these places are marked by a person making a call.  However, there are other places that a person travels through on a road route in which they did not make a call.  The logic behind this assumption is fundamentally that of a missing data problem: we do not know how long a person stayed in each place, how important was each place to a particular person, or if places where they made calls were more or less important than other places they traveled through.  Consequently, these measures assume all places along a person's road route are of equal importance and counts them all. 

To calculate the group B measures, we take every pair of sites that are consecutive in a spatiotemporal trajectory $M$ and identify the grid cells that belong to the quickest road route between the two sites. We form the set of all the grid cells $\mathcal{V}(M)$ on these quickest road routes which also include their start and end cells, the sites from which calls were made. A grid cell appears only once in $\mathcal{V}(M)$. The elements of $\mathcal{V}(M)$ are called visited grid cells. The visited grid cells that have cellular towers in them are called visited sites.

{\bf Grid cells visited (GCV-R)}. This measure is given by the number of visited grid cells. The GCV-R measure for $\mathcal{P}$ is equal to 93 (98th percentile) because there are 93 visited grid cells between sites 1976 and 360. This measure relies on the assumptions that a person must have existed on the ground in places between subsequent calls and that, without further information, all places a person might have visited are equally important.

{\bf Sites visited (SV-R)}. This measure is the ratio between the number of visited sites and the total number of sites in the reference time period of the trajectory $M$. As discussed in {\bf SI Appendix, Section SI2}, the number of sites could change from a reference time period to another as cellular towers are installed or decommissioned. Thus adjusting for the time varying number of sites is required to define a measure whose values are consistent across reference time periods. For example, 19 out of the 93 visited grid cells between sites 1976 and 360 were sites in October 2005. Since the total number of sites in October 2005 is 53, the measure SV-R is $19/53=0.358$ (98th percentile) for $\mathcal{P}$. The definition of the SV-R measure is based on the assumption that there is something important about where a cellular tower is placed, either high population density, high through-traffic, near an important area, at a cross-roads, etc.  The reason that cellular towers are located in certain areas might differ between contexts and across time, but what does not differ is that there is likely a reason for cellular tower placement.  We use this particular assumption because it assumes the least possible in order to define a place and is therefore the most comparable across contexts and time. 

These two measures take into account the spatial range of a person's mobility: the further each trip, the larger the number of sites and grid cells visited. But the frequency of movement is not captured by these measures. If $\mathcal{P}$ would make a third call from site 1976, a fourth call from site 360, a fifth call from site 1976, and so on, the values of GCV-R and SV-R will remain the same.

\subsection{Group C: Measures of spatial range and frequency of mobility} 

The final three measures of mobility calculate the sum of distances between sites where consecutive communication episodes occurred. They differ only in terms of the type of units of distance used. These distances are related to the quickest of all possible road routes between two sites that are consecutive in a spatiotemporal trajectory.

{\bf Distance traveled (DT-R)}. The distance metric for this measure is the length of the quickest road route. There are two key differences between DT-R and the existing measure DT-SL: (i) DT-SL involves movement between cellular towers, while DT-R involves movement between sites; and (ii) DT-SL is the sum of straight line distances, while DT-R is the sum of distances via road travel. If two consecutive calls were made using two towers that belong to the same site, DT-SL will record the straight line distance between the two towers while DT-R will not record any movement. On the other hand, DT-SL will underestimate distances between two points since straight line distances are almost always (if not always) smaller than distances via roads. The DT-R measure of $\mathcal{P}$ is 432.1 km (38th percentile) since this is the length of the quickest road route between the two sites from which $\mathcal{P}$ called.

{\bf Time traveled (TT-R)}.  The distance metric for this measure is the estimated travel time on the quickest road route. Travel time can be estimated in several ways.  Speed limits can be used where available.  If speed limit information is not available or the quality of roads is such that speed limits cannot be met, then average travel speeds must estimated for each type of road  \--- see {\bf SI Appendix, Section SI1.1}.  The TT-R of $\mathcal{P}$ is 6 hours and 4 minutes (33rd percentile) since this is the smallest estimated travel time via roads between the two sites from which calls were made.

{\bf Grid cells traveled (GCT-R)}. The distance metric for this measure is the number of grid cells intersected by the quickest road route. The start site of a route is counted, but the end site is not counted. Sites that are both the end of one route and the beginning of another are not counted twice. The GCT-R of $\mathcal{P}$ is 92 (42nd percentile) since there are 93 grid cells on the quickest road route between the two sites from which $\mathcal{P}$ called, including the start and end sites.

These three measures all incorporate the frequency and the spatial range of a person's mobility.  The more times a person moves, the higher will be their distance, time, and grid cells traveled.  The further each trip, the higher will be these measures as well. When jointly employed, these three distance metrics are useful in identifying various patterns of mobility. For example, the mobility of two individuals that travel the same distance but use different types of roads (e.g., highways vs. minor country roads) will be rated the same by DT-R, but will differ with respect to TT-R. The distance metric for GCT-R is less dependent on particular shapes of the roads or their quality which might help when comparing mobility for spatial trajectories recorded in distant regions or countries.

\section{Assessment of the Proposed Measures of Mobility}
The assessment of CDR mobility measures is limited by the reality that there currently exists no standard measure of mobility, or no gold standard to which we can compare new measures. In this regard, the most important assessment tool available is face validity.  In other words, the best assessment tool is a careful discussion of which measures make sense and if they actually measure what we think they should be measuring. Part of this face validity discussion is above in the description of the measures, dimensions of mobility, and assumptions required for each measure. 

We undertake additional assessment of our six new measures against the existing measures of mobility and against each other by estimating longitudinal pairwise correlations based on the spatiotemporal trajectories of callers for each of 44 months of Rwandan CDRs. Results, figures and a discussion are presented in {\bf SI Appendix, Section SI5}. All longitudinal associations are positive with values from medium to high, and are very stable through time. Measures within groups have the strongest associations, as we would expect. But other, less intuitive high associations emerge, especially between certain existing and new measures. In particular, the DT-SL measure, which is conceptually consistent with both key dimensions of mobility, has the strongest associations with the measures in group B (spatial range), and only the second strongest associations with the measures in group C (spatial range and frequency). While this is somewhat surprising, it also emphasizes the fundamental differences in the way DT-SL is defined as opposed to our new measures, especially DT-R.

In addition to face validity and correlations, it is important to assess which groups of measures and which measures within each group should be used for studies of population mobility. The six measures we introduce offer multiple choices of combinations which could be selected based on particular research questions and contexts. We also argue that all our six measures are needed in principled, thorough population mobility studies. Despite having common characteristics in terms of the two key dimensions of mobility we discussed, each measure captures a slightly different aspect of mobility, and is thus relevant alone as well as jointly with the other five measures. To demonstrate this, we used our six measures to define categories of callers with different mobility profiles for four of the 44 months of data (June 2005 to January 2009) \---- see {\bf SI Appendix, Section SI6}. A monthly spatiotemporal trajectory of a caller was classified as having high or low mobility with respect to a measure if the value of their mobility measure was above or below the median of observed values during that month. For each month, with six separate measures this leads to 64 categories. Tables S1\---S4 ({\bf SI Appendix}) show that at least 11 of the 64 categories contain at least $1.0\%$ of the callers in each of the four  months we examined. This suggests that there are many distinct mobility types in this population that can only be identified by using all six measures in combination. Overlooking any one measure would lead to conflating segments of the population with distinct mobility profiles. 

Further, we find several notable patterns with these tables. Two profiles which rate mobility as low or high for all six measures are the largest ones in all four months, and comprise about 30\% of the monthly callers. The third and fourth largest segments rate mobility as low or high for groups A and C, and as the opposite for group B. These segments comprise about 7\% of the monthly callers, and show the relevance of capturing frequency of mobility (groups A and C) versus spatial range (group B). Another notable result is the relatively common mobility profiles that rank high on one spatial range measure, low on the other spatial range measure, and high or low on all other measures. These two groups comprise about 5\% of the population and indicate people who likely travel often and far, but mostly in areas with few cellular towers (thus a high GCV-R but low SV-R). In contrast, there are people who travel seldom and short distances, but in areas with many cellular towers (thus a low GCV-R but high SV-R). Again, assessing all six mobility measures for each individual is necessary to identify particular mobility types in a population.

The mobility measure time traveled (TT-R) has an additional important use. As we show in {\bf SI Appendix, Section SI7}, the values of this measure can be used to identify and possibly filter out spatiotemporal trajectories affected by errors in cellular services provider's databases, or by intruders who gain unauthorized access to mobile phones and use them to communicate at the same time as the actual owners. The identification of such unusual trajectories is not possible with the existing measures of mobility.

\section{Discussion}
Our mobility measures are designed to be applicable to any research setting, from wealthy countries with well-developed mobile phone and transportation infrastructure, to poorer countries that are yet developing transportation and communication networks.  They constitute an important advance in the social scientific study of mobility which could lead to improved understanding of human health and well-being and macro-economic, social, political, and demographic dynamics. Being almost entirely spatially derived, and using CDRs enhanced with GIS data, these new measures circumvent many of the problems inherent in existing mobility measures and are independent of cellular tower density and the social, political, economic, or demographic characteristics that influence tower density. They are thus relevant and comparable in different contexts.  Another key goal with this paper is to stimulate discussions on mobility measures using CDRs, and to promote social science research on the causes and consequences of human mobility. In this regard, we herewith discuss some of the many ways in which these new CDR-based measures of mobility can be used to enhance and expand our understanding of human well-being and social organization.

First, these new measures can replace older measures, often based on sample surveys, to improve understanding of existing mobility related questions.  The benefit here is clear, given that CDR-based measures can significantly increase the accuracy, detail, and time period over which mobility can be recorded.  They are also much less costly to obtain than detailed survey measurements.  CDRs can be collected and measures of mobility calculated for respondents who participate in sample surveys, giving the researcher not only immense detail about respondent mobility, but the opportunity to compare it with survey records of other characteristics and behaviors. 

Second, these new measures open up entirely new avenues of research.  Because CDRs can cover millions of people, it is possible to calculate population-level mobility measures.  For example, one can calculate a measure of general mobility for a city, state, province, or region.  This could then be compared to individual level behaviors and outcomes to investigate questions such as how population mobility influences individual migration, tuberculosis infection, or women`s labor force participation. Population level mobility can also be related to population-level characteristics, such as HIV prevalence rates, birth rates, social norms, economic well-being, or political participation.  With sample surveys, it has never before been possible to calculate population level characteristics, thus CDR-based measures, if appropriately calculated to be independent of tower density and the related contextual characteristics, create new and possibly groundbreaking opportunities for social science.

Third, CDR-derived population level measures mobility can be used to identify and study emergency events, such as natural disasters and armed conflict.  For example, theory and evidence predict that people will change their mobility patterns during and after an earthquake or a large bomb blast \cite{bagrow-et-al-2013,lu-et-al-2012}.  With access to real time data, it would then be possible to pinpoint an earthquake or bomb blast in real time, even in remote areas with poor communication and transportation linkages.  Given that the time to humanitarian response significantly influences the magnitude and time of the post-event disaster period, real-time identification of hazardous events could ultimately lead to decreasing the human toll of disasters.

While CDR-based measures can create immense new opportunities for understanding human mobility, there are several limitations of which researchers must be aware.  As with all organic big data \cite{Groves-2011}, selection is a major concern.  For mobile phone data, mobile phone users are included in the data set and non-users are excluded.  Research suggests that users are more likely to be male, educated, and live in urban areas \cite{Weslowski-et-al-2013b,Blumenstock-Eagle-2012}.  Alternately, research has also shown that there are an estimated 90.2 mobile phones per 100 people in poorer countries \cite{ITU-2014}.  Considering that mobile phone penetration statistics are largely analogous to response rates in surveys, we can say that CDR based data essentially have a 90.2\% response rate in poorer countries, which is generally considered good if not excellent, regardless of selection.

Another key limitation to the use of CDR-based mobility measures is the inherent error.  The primary problem is that although mobile calls are recorded as occurring at a cellular tower, the person making the call is rarely at that tower.  Instead they are likely to be within 5 or 10 km from the tower, depending on the type of antenna used in the tower and topography.  However, we argue that the benefits of CDR-based mobility measures vastly outweigh the detractions, especially when compared to the alternative of survey-based measures with inherent error due to human difficulties in recalling location, time, and movement accurately and the inability to measure population level mobility.

\section*{Acknowledgments}
The authors thank Daniel Bjorkegren and Joshua Blumenstock for their help with the initial stages of processing of the call data records. This work was partially supported by National Science Foundation Grant DMS-1120255 (to A.D.). N.W.'s contribution has benefited from generous support of a NIH Pathways to Independence grant (R00HD067587).

\appendix

\setcounter{figure}{0}
\renewcommand{\figurename}{Figure }
\renewcommand{\thefigure}{S\arabic{figure}}

\renewcommand{\tablename}{Table }
\renewcommand{\thetable}{S\arabic{table}}

\section*{Supplementary Information (SI) Appendix}

\subsection*{SI1: The Road Network and the Grid Cell System}

Two of our methodological developments that are based on Geographic Information Systems (GIS) are the road network system from Rwanda and the grid cell system which divides a spatial bounding box for Rwanda's boundary into 2040 5 km x 5 km cells. With any GIS, the choice of an appropriate coordinate system, determined by the geographic location and scale of analysis, is required to assure accurate measures of distance and area. Our road network and grid cell system are based in the Universal Transverse Mercator (UTM) zone 36S coordinate system, using the WGS84 datum\footnote{Such types of linear and areal measurements cannot be performed in a GIS using an unprojected or spherical coordinate system, such as degrees latitude and longitude, which measure angles rather than surface distance.}. 

\begin{figure}[htbp!]
\centering
 \includegraphics[height=4in,angle=0]{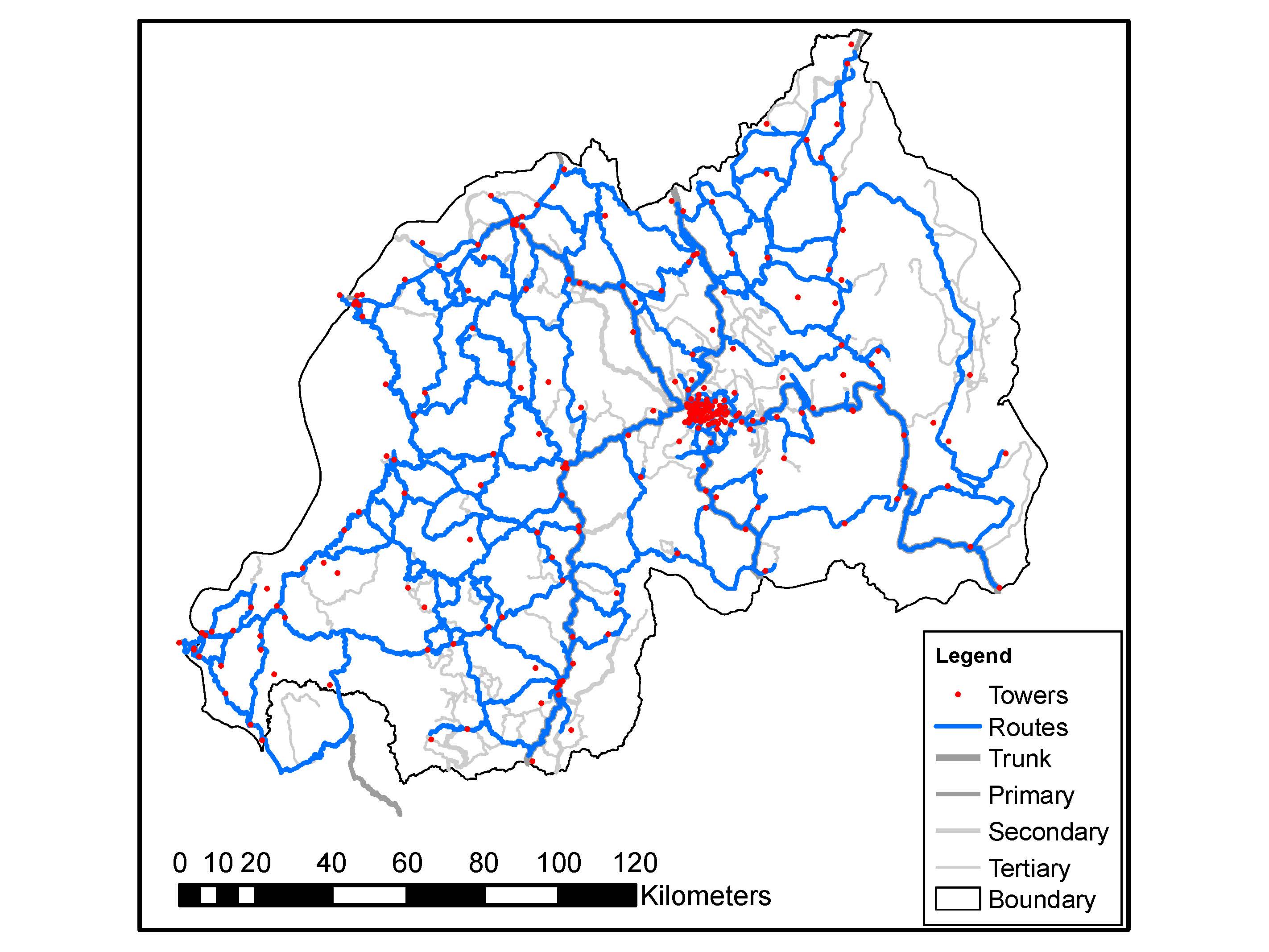}
 \caption{\label{fig:towersRoads} Map of Rwanda showing the position of the 269 cellular towers (\textcolor{red}{red}) and the structure of the network of roads (trunk, primary, secondary and tertiary) used for our mobility measures (\textcolor{gray}{gray}). Roads that are also segments on quickest routes are shown in \textcolor{blue}{blue}.}
\end{figure}

\begin{figure}[htbp!]
\centering
    \includegraphics[height=4in,angle=0]{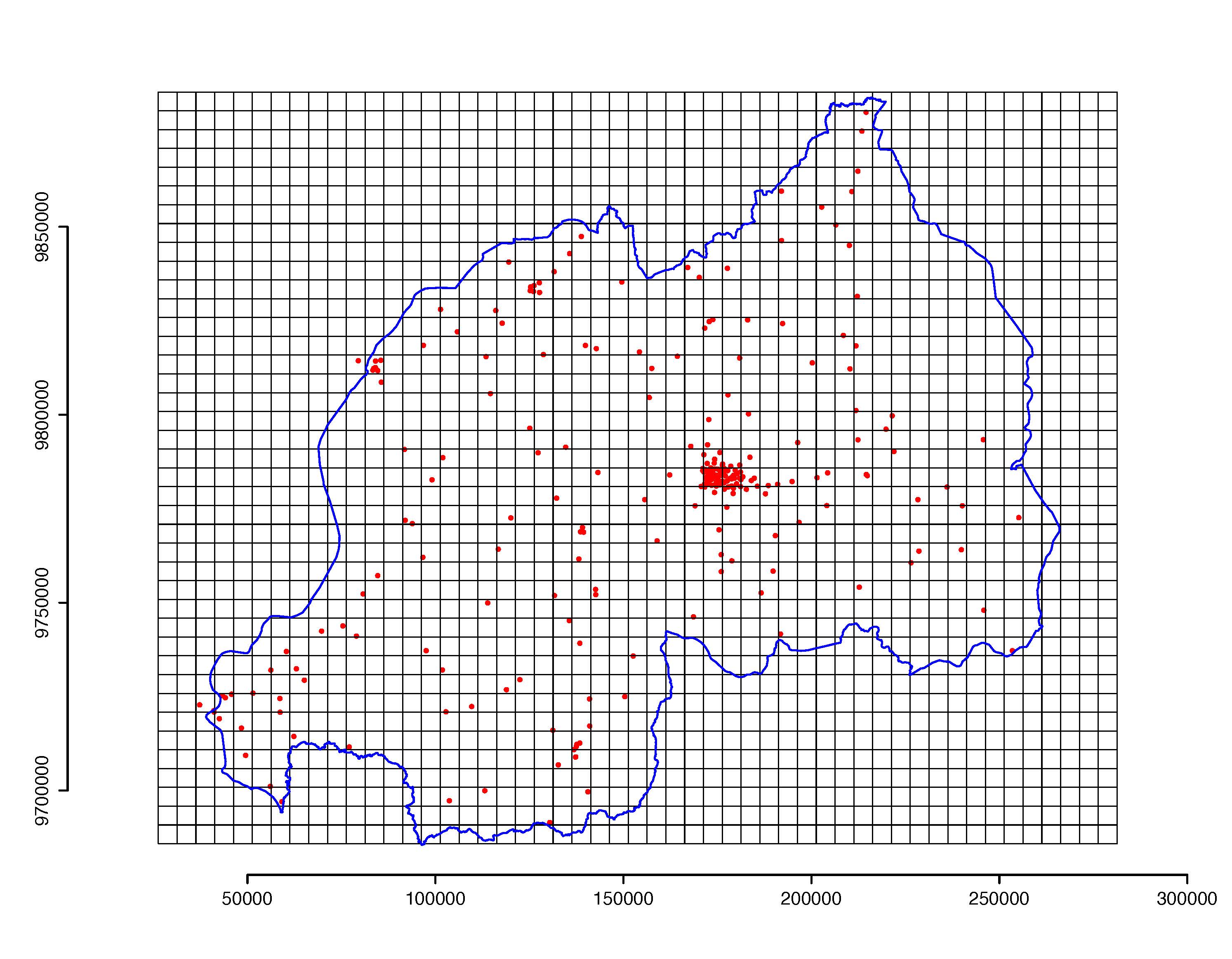}
 \caption{\label{fig:towersGridcells}Map of Rwanda showing the position of the cellular towers (\textcolor{red}{red}) with respect to the 2040 5 km x 5 km grid cells. Rwanda's boundary is shown in \textcolor{blue}{blue}.}
\end{figure}

Figures \ref{fig:towersRoads} and \ref{fig:towersGridcells} display the locations of the $269$ cellular towers that appear in the Rwandan CDR data with respect to the road network and the grid cell system, respectively. The grid cells that contain at least one tower are called sites. Only 155 out of the 2040 grid cells are sites. Four sites in the Kigali area contain the largest number of cellular towers: 41, 22, 6 and 5, respectively. Seven sites contain four towers, four sites contain three towers, 14 sites contain two towers and the other sites contain only one tower. These counts represent the towers that belong to a site between June 1, 2005 and January 31, 2009. In any period of time between these dates, all, some or none of the towers that belong to a site are actually active (i.e. handle cellular communications). As such, the number of sites (i.e., grid cells that contain at least one active tower) in a time period might be smaller than 155 \--- see Section SI2.

\subsubsection*{SI1.2: The Grid Cell System}

Our approach, which involves overlaying a customized rectangular grid with square cells of equal size on the map of Rwanda, creates a systematic method of circumventing the major problem of spatial variance in cellular tower density.  We replace tower locations with the centroid of the sites they belong to. Instead of measuring straight line distances from tower to tower, we measure distances between the centroids of the sites via the quickest road route which connects these centroids. 

Choosing the size of the grid cells is an important decision. We are interested in country level mobility patterns and believe that 5 km x 5 km grid cells are appropriate for this purpose\footnote{For other applications that involve an in-depth look at mobility within predominantly urban settings such as Rwanda's capital Kigali, the same methodology can be used with the grid cell size modified to a more appropriate scale to capture local scale mobility (e.g., 1 km x 1 km grid cells).}. Based on geographical and technological considerations, we estimated catchment areas in which a user of a cellular tower is likely to be located. We estimated that the maximum signal distance for the type of towers in Rwanda is roughly 10 km. Several factors further reduce this maximum signal distance, including relative location of a user with respect to a tower, topography of the areas surrounding towers, and the decay in signal strength with increasing distances from towers. As such, we  reduced the maximum user-to-tower distance to 5 km. The resulting 5 km x 5 km grid cell system is a 51 x 40 matrix (2040 grid cells) that covers 51,000 km$^2$ extending just outside of the border of Rwanda \--- see Figure \ref{fig:towersGridcells}. Each grid cell is indexed by a number from 1 to 2040: grid cell 1 is located in the lower left corner and grid cell 2040 is located in the upper right corner. The indices increase first by row, then by column. Each of the 269 towers is subsequently mapped to its corresponding grid cell (site).  

A possible caveat of using a grid cell system as a foundation for measuring mobility is that the spatial placement of the grid creates arbitrary boundaries that could non-systematically influence mobility measures.  For example, person $\mathcal{A}$ could call from a cellular tower that is one meter away from a grid cell boundary.  If $\mathcal{A}$ moves two meters to cross the grid cell boundary and makes a call from a tower in the new grid cell, $\mathcal{A}$ will be registered as moving between grid cells, even with only two meters of actual travel.  On the other hand, another person $\mathcal{B}$ could call from a tower inside a grid cell, move up to $5\sqrt{2}\approx 7.07$ km without crossing a grid cell boundary, and make another call. Because $\mathcal{B}$ remained in the same grid cell, $\mathcal{B}$ would not be registered as having moved, despite traveling a lot further than $\mathcal{A}$.

There is a reasonably simple, though computationally intensive, method to account for the bias induced by the arbitrary spatial placement of the grid cells. Consider a system of 5 km x 5 km grid cells which is placed over a map of the study area, and calculate the corresponding values of the measures of mobility which are dependent on the locations of the cells.  Next, move the entire grid system 1 km east over the map, and recalculate the values of the mobility measures. The grid system can be moved 25 times (by 1 km east and 1 km south each time), and the resulting 25 sets of values of the measures of mobility can be determined.  These 25 sets of values can be used to produce mobility measures estimates as well as standard errors which account for the arbitrary placement of grid cell boundaries.

The raw OSM's data for Rwanda was such that 11 sites were not intersected by the Rwandan road network. To connect these sites to the road network, we moved the location of their centroids to adjacent grid cell centroids. The adjacent grid cell used was the one closest to the tower in the original cell. In a few cases this led to an overlap between two sites, one of which happened in the Kigali area which has a high tower density. We believe this is not a major problem because, as mentioned above, the coverage of a cellular tower is roughly a circular area with a 5 km radius. The distance between two centroids is exactly 5 kilometers and each of the roads that were barely missed was within 3 km of the centroid of the nearest grid cell. In all cases, we spot checked the change to make sure it was the most reasonable. 

When we replace towers with grid cells, there is the possibility of increasing the error in a person's location. Notably, there is the symmetric possibility that the grid system could decrease locational error. In most cases, it is likely that the combined error (uncertainty of a person's location in relation to a tower combined with additional grid system locational error) is negligible.  In the most extreme circumstance, with a 5 km x 5 km grid system, towers that broadcast to a 10 km radius, and a tower that is in a far corner of a grid cell, a person's location could be calculated as being up to 13.5 km  from their actual location.  Note that the majority of the error here (10 out of 13.5 km) is due to tower-location uncertainty and the minority of error (3.5 km) is due to the imposition of the grid system.  This maximum possible error of 13.5 km is likely not as problematic when measuring mobility on a national scale compared to a smaller local scale.  When measuring mobility on a smaller scale in areas with higher tower density, or towers that are closer to each other than 10 km, the maximum possible locational error will be less.  It can be further reduced by decreasing the size of grid cells.  Thus, locational error must be carefully considered when CDR-based data is used to measure mobility and further work should be done to assess the effects of selection and locational error.  

\subsection*{SI2: The Temporal Dynamics of the Cellular Network}

The network of cellular towers managed by a wireless service provider could vary significantly over months  and years in terms of the total number of towers, their spatial coverage, and the number of users of the network. A cellular tower is called active in a given time period (e.g., a month or a year) if it handled at least one call during that period. We used the available Rwandan CDRs to determine the number of callers in this provider's network, and which cellular towers handled their calls every month from June 2005 to January 2009.

\begin{figure}[htbp!]
\centering
    \includegraphics[height=2.8in,angle=0]{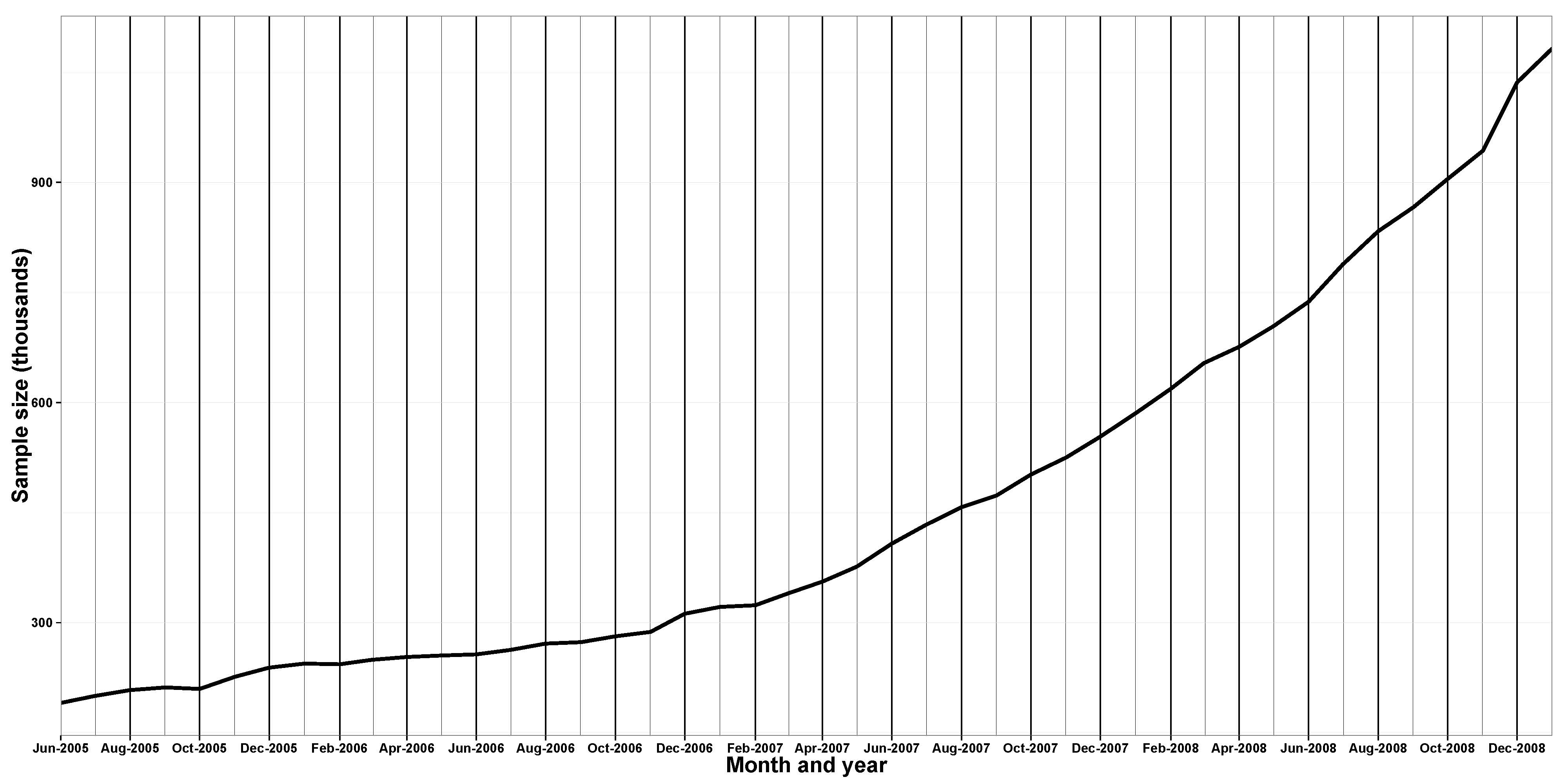}
\caption{\label{fig:sampleSizeCallers}Number of callers during each month between June 2005 and January 2009.}
\end{figure}

\begin{figure}[htbp!]
\centering    
  \includegraphics[height=2.8in,angle=0]{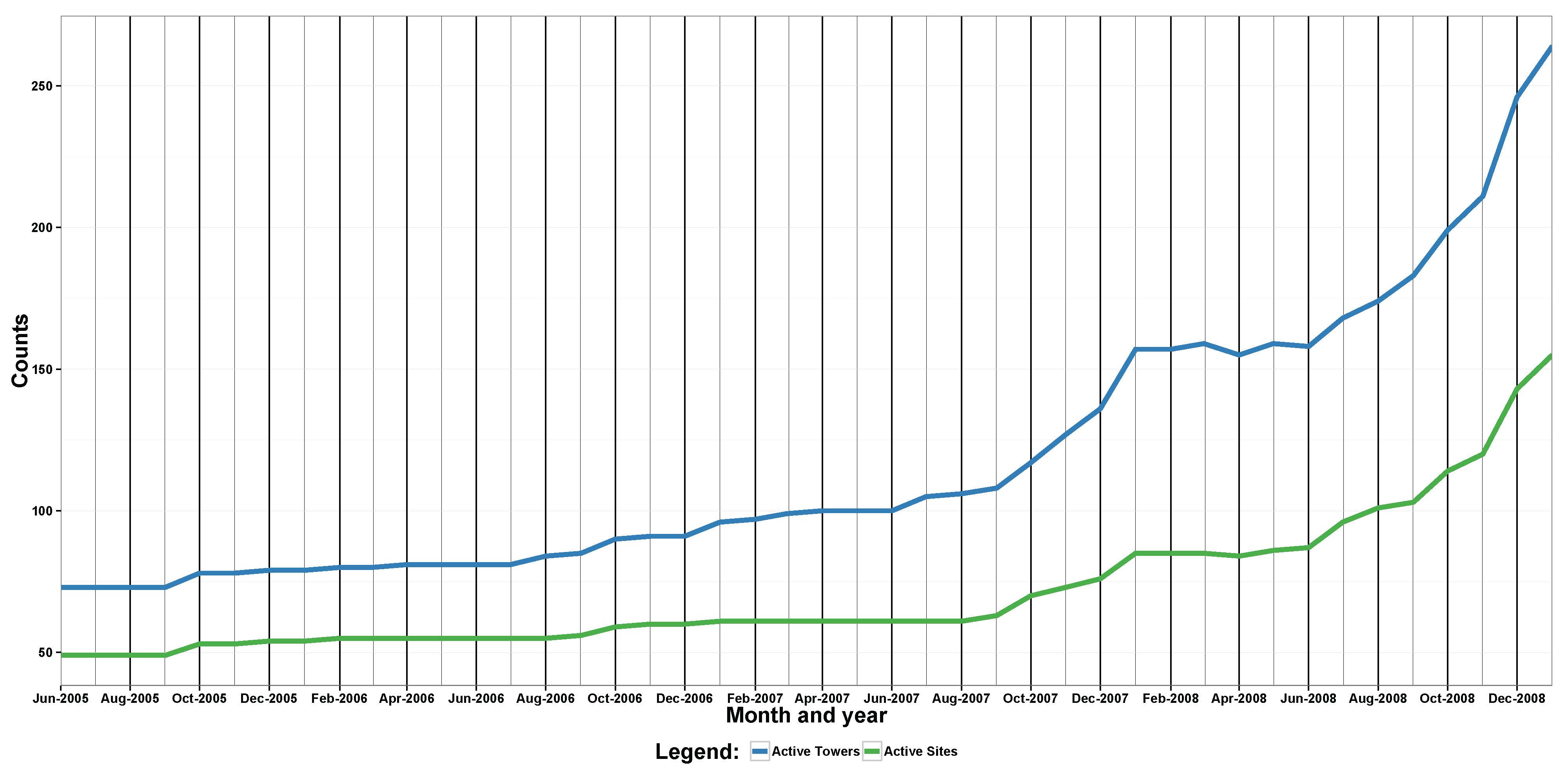}
\caption{\label{fig:activeTowersSites}Number of active cellular towers (\textcolor{blue}{blue}) and sites (\textcolor{green}{green}) during each month between June 2005 and January 2009.}
\end{figure}

\indent Figure \ref{fig:sampleSizeCallers} shows that the number of users using this cellular network increased from 190 thousands in June 2005 to 238 thousands in December  2005, 310 thousands in December 2006, 552 thousands in December 2007, and reached more than 1 million people by December 2008. Figure \ref{fig:activeTowersSites} reveals that the number of active towers that handled communications in this network continually increased from 73 in June 2005 to 79 in December 2005, 91 in December 2006, 136 in December 2007 and reached 246 in December 2008. During this time the network expanded with progressively increasing tower density in more populated areas through the installation of additional towers in sites that already had towers in them, but also with progressively increasing spatial coverage through the installation of new towers in grid cells that previously did not contain any towers. The difference between these two dimensions of the dynamics of the cellular network is evident when comparing the month to month increase in the number of sites with the increase in the number of towers: increased tower density at the same sites is not captured in the number of sites during each month, but it is captured in the increase in the number of active towers. Although not obvious in this figure, there is always the possibility that a cellular network loses towers which leads to lower tower density at some sites or to sites no longer containing active towers. Increased spatial coverage is evident from the bottom panel of Figure \ref{fig:activeTowersSites}: there were 49 sites in June 2005, 54 sites in December 2005, 60 sites in December 2006, 76 sites in December 2007 and 143 sites in December 2008.

\begin{figure}[htbp!]
\begin{center}
\includegraphics[height=4in,angle=0]{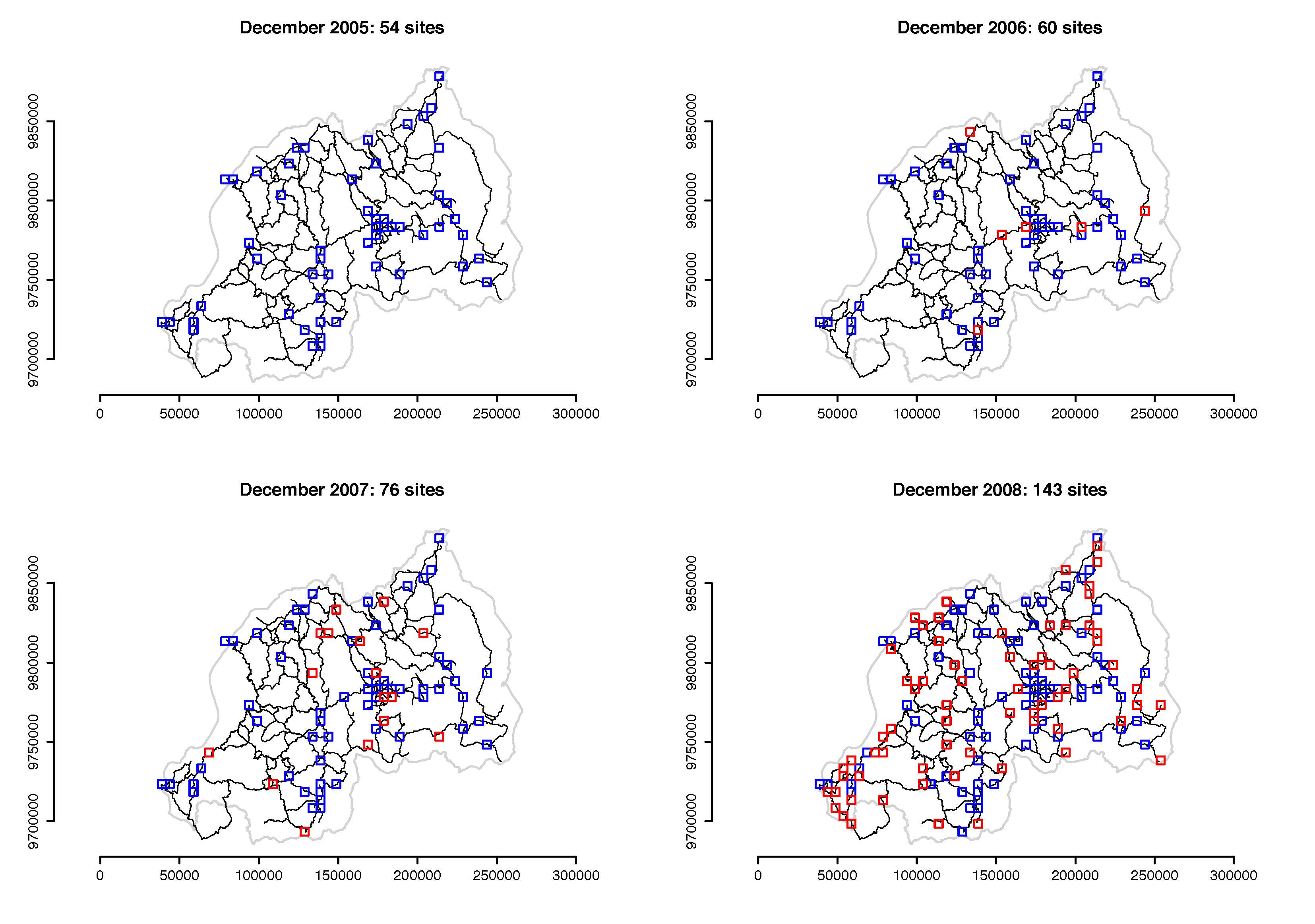}
\caption{Location of sites during four months: December 2005, 2006, 2007 and 2008. In each of the four panels, \textcolor{red}{red} denotes grid cells that were not sites the year before. All the other sites are shown in \textcolor{blue}{blue}. This plot reveals that a large number of cellular towers were installed between December 2007 and December 2008 which significantly increased the spatial coverage of the wireless services provider that provided the Rwandan CDRs.} 
\label{fig:placesDynamics}
\end{center}
\end{figure}

\indent The expansion of the spatial coverage of this cellular network is shown in Figure \ref{fig:placesDynamics}, which shows where the installation of new towers led to coverage in grid cells that previously did not have cellular service from this provider. The large expansion in spatial coverage recorded in December 2008 compared to December 2007 is especially important as it vastly improved the accessibility of mobile technology in towns and rural areas throughout Rwanda.

\subsection*{SI3: Measures of Mobility}

Here we give formal mathematical definitions of the measures of mobility described in the main text. Consider the sequence of CDRs associated with a mobile phone in a reference period of time $\mathcal{T}$ (e.g., a day, a week, a month or a year):
\begin{eqnarray} 
 M = \{ m_1,m_2,\ldots,m_n\}.\label{eq:cdrSeq}
\end{eqnarray}
We assume that the wireless-service provider that generated these CDRs has $K$ active towers in the reference time period $\mathcal{T}$, and that the spatial locations $l^{CT}_i$, $i \in \mathcal{K}=\{1,2,\ldots,K\}$ of these active towers are known. In (\ref{eq:cdrSeq}), $m_i\in \mathcal{K}$, $1\le i\le n$, is the identifier of the cellular tower that handled the communication represented by the $i$-th CDR in the sequence (\ref{eq:cdrSeq}). If $i<j$ the communication represented by $m_i$ was recorded before the communication represented by $m_j$.  We refer to $M$ as the spatiotemporal trajectory of the cellular phone that generated the sequence of CDRs. We remark that more than one tower might have handled the same communication (e.g., a call), but in that case several CDRs \--- one for each cellular tower \--- would have been generated. 

For any pair of spatial locations $l$ and $l^{\prime}$ identified by their latitude and longitude coordinates, we define the distance function $d_{SL}(l,l^{\prime})$ which represents the straight line or ``as the crow flies'' distance between $l$ and $l^{\prime}$. We take $d_{SL}(l,l)=0$.

\subsubsection*{SI3.1: Existing Measures of Mobility}

For the spatiotemporal trajectory $M$ from (\ref{eq:cdrSeq}), the measure of mobility called ``number of towers used" (NTU) is the number of unique towers that appears in this sequence, i.e.

$$
 \#\left\{i: i\in \mathcal{K} \mbox{ such that there exists }m_j,1\le j\le n \mbox{ with } m_j=i\right\}.
$$
Here $\#A$ denotes the number of elements in the set $A$. The measure of mobility called ``distance traveled" (DT-SL) is the sum of straight line or ``as the crow flies'' distances between consecutive towers from which communication occurred:
$$
 \sum\limits_{j=2}^n d_{SL}\left(l^{CT}_{m_{j-1}},l^{CT}_{m_{j}}\right).
$$
The measure of mobility called ``maximum distance traveled" (MDT) is the maximum straight line distance between two towers in the sequence $M$:
$$
 \max\limits_{1\le i<j\le n} d_{SL}\left(l^{CT}_{m_i},l^{CT}_{m_{j}}\right).
$$
The measure of mobility called ``radius of gyration'' (RoG) is the square root of the mean of the squared straight line distances between the locations of towers in $M$ and the center of mass 
\begin{eqnarray}
\bar{l}_M=\frac{1}{n}\sum\limits_{j=1}^n l^T_{m_j}, \label{eq:centermass}
\end{eqnarray}
\noindent of the trajectory:
$$
 \sqrt{\frac{1}{n}\sum\limits_{j=1}^n d^2_{SL}\left(l^T_{m_j},\bar{l}_M\right)}.
$$
\noindent Equation (\ref{eq:centermass}) defines the center of mass as the arithmetic mean of the spatial locations of cellular towers in the trajectory $M$. 

\subsubsection*{SI3.2: New Measures of Mobility}

We assume that the region of interest was divided into non-overlapping grid cells identified by indices in $\mathcal{Q}\in \{1,2,\ldots,Q\}$. We denote by $l^{GC}_j$ the location of the centroid of the grid cell $j\in \mathcal{Q}$. We introduce a mapping function $q^{GC}(\cdot)$ which gives, for each cellular tower $i\in \mathcal{K}$, the grid cell $q^{GC}(i)\in \mathcal{Q}$ the tower belongs to. The sites are those grid cells that contain at least one tower:
$$
 \mathcal{S} = \left\{j: j\in \mathcal{Q} \mbox{ such that there exists }  i\in \mathcal{K} \mbox{ with } q^{GC}(i)=j\right\}.
$$
Since we assume that all the towers indexed by $\mathcal{K}$ are active in the reference time period $\mathcal{T}$, $\mathcal{S}$ represents the set of sites in $\mathcal{T}$.

We transform the spatiotemporal trajectory $M$ from (\ref{eq:cdrSeq}) into the corresponding time ordered sequence of sites to which the active towers that appear in $M$ belong to:
\begin{eqnarray}
 M^{GC} = \{ g_1,g_2,\ldots,g_n\}, \label{eq:gridSeq}
\end{eqnarray}
\noindent where $g_i = q^{GC}(m_i)\in \mathcal{S}$.

The measure of mobility called ``number of trips" (NT) is a count of the number of times a person communicates from a different grid cell than their previous communication:
$$
 \# \left\{ i: i\in \{1,2,\ldots,n-1\} \mbox{ such that } g_i\ne g_{i+1}\right\}.
$$

The other five measures of mobility introduced in the main text are constructed with respect to a road network that connects any two locations $l$ and $l^{\prime}$ in the region of interest. If  a location $l$ is not on the road network, we assume its spatial location is replaced by its projection (the location at the smallest straight line distance) on the road network. With this convention, we assume that any two locations $l$ and $l^{\prime}$ are connected by at least one continuous subset of locations called road route on the road network. We assume that these road routes do not contain loops. From all the road routes that connect $l$ and $l^{\prime}$, we choose the route $\mathcal{P}(l,l^{\prime})$ that is quickest, i.e. the route with the smallest estimated travel time. Each location on the path $\mathcal{P}(l,l^{\prime})$ must belong to exactly one grid cell. We denote by $\mathcal{R}(l,l^{\prime})$ the subset of $\mathcal{Q}$ that comprises all the grid cells intersected by the road route $\mathcal{P}(l,l^{\prime})$. We define the following distance functions:
\begin{enumerate}
 \item $d_{RD}(l,l^{\prime})$ represents the length of $\mathcal{P}(l,l^{\prime})$.
 \item $d_{TT}(l,l^{\prime})$ represents the estimated travel time between $l$ and $l^{\prime}$ on $\mathcal{P}(l,l^{\prime})$.
 \item $d_{GC}(l,l^{\prime})$ represents the number of grid cells in $\mathcal{R}(l,l^{\prime})$.
\end{enumerate}
We take $d_{GC}(l,l)=d_{RD}(l,l)=d_{TT}(l,l)=0$.

The set of visited grid cells associated with the spatiotemporal trajectory $M^{GC}$ from (\ref{eq:gridSeq}) are those grid cells on the fastest road routes that connect consecutive sites in $M^{GC}$:
$$
 \mathcal{V}\left( M^{GC}\right) = \bigcup\limits_{j=2}^{n} \mathcal{P}\left( l^{GC}_{g_{j-1}},l^{GC}_{g_{j}}\right).
$$

The measure of mobility called ``grid cells visited" (GCV-R) is given by the number of visited grid cells $\# \mathcal{V}\left( M^{GC}\right)$. The visited grid cells that are also sites are called visited sites. The measure of mobility called ``sites visited" (SV-R) is given by the ratio between the number of visited sites and the total number of sites in the reference time period $\mathcal{T}$:
$$
 \#\left( \mathcal{V}\left( M^{GC}\right) \cap \mathcal{S}\right)/\#\mathcal{S}.
$$
The measure of mobility called ``distance traveled" (DT-R) is the sum of the lengths of the quickest road routes between consecutive sites from which communication has occurred:
$$
 \sum\limits_{j=2}^{n} d_{RD}\left( l^{GC}_{g_{j-1}},l^{GC}_{g_{j}}\right).
$$
The measure of mobility called ``time traveled" (TT-R) is the sum of the estimated travel times on the quickest road routes between consecutive sites from which communication has occurred:
$$
 \sum\limits_{j=2}^{n} d_{TT}\left( l^{GC}_{g_{j-1}},l^{GC}_{g_{j}}\right).
$$
The measure of mobility called ``grid cells traveled" (GCT-R) is the number of grid cells intersected by the quickest road routes between consecutive sites from which communication has occurred:
$$
 \sum\limits_{j=2}^{n} d_{GC}\left( l^{GC}_{g_{j-1}},l^{GC}_{g_{j}}\right).
$$

\subsection*{SI4: Addressing the Possibility of Air Travel for the Proposed Mobility Measures}

Our proposed system of calculating mobility measures is entirely based on the assumption that most people travel via land and on roads.  Our empirical work employs CDRs from Rwanda which is a small country of approximately 23,338 km$^2$ \---- about the size of the U.S. state of Massachusetts.  Rwanda generally ranks low on indices of human development, with an estimated 57\% of the population poor and 38\% extremely poor in 2006.  The population density is amongst the highest in Africa, at 347 people per square km and most Rwandans (87\%) live in rural areas. Despite the rather poor and rural conditions of the population and contentious history, the Rwandan economy, road networks, and living conditions of the people have improved throughout the 2000Õs.

As such, the assumption that most travel occurs by roads is quite reasonable for a country such as Rwanda. However, in larger and/or wealthier countries a significant proportion of travel occurs by air instead of roads. Several adjustments must be made to our methods to account for air travel. The first step in making these adjustments is to identify when a person most probably traveled via air rather than roads.  To this end, we must make explicit use of the dates and the times of generation of each CDR. If the time period between two consecutive CDRs of a person $\mathcal{A}$ associated with different grid cells $G_1$ and $G_2$ is shorter than the shortest possible travel time via roads between these grid cells, we can assume that $\mathcal{A}$ traveled via air between $G_1$ and $G_2$.  As with any assumption, this inherently includes some error.  This method will correctly identify all air travel for people who make calls soon before taking off and soon after landing.  However, if a person does not make calls for some time before and after flying, then their travel will not be identified as such.  There will be more error in this manner for flights that cover shorter distances than for longer distance flights.

Once air travel movement has been identified for pairs of consecutive CDRs, four of the six proposed mobility measures could be calculated differently. Specifically, for such pairs, sites visited (SV-R) can include only the grid cells in which $\mathcal{A}$ made consecutive calls and no other sites (grid cells with active mobile towers) on a route in between. Grid cells visited (GCV-R) can be the number of sites visited and no other grid cells on a route in between. Time traveled (TT-R) can be the difference between the times of the generation of the two consecutive CDRs. Alternately, the time that flights take between the two closest airports can be used. Distance traveled (DT-R) can be calculated as the straight line distance between the centroids of the two sites from which calls were made.  However, to maintain consistency with segments of movement that most likely occurred by roads, it might be desirable to  calculate all air travel based measures as if they were undertaken via road.  In either case, the number of trips (NT) requires no adjustment for air travel movement.

\subsection*{SI5: Longitudinal Pairwise Associations of Measures of Mobility}

We investigate the relationships between the existing measures of mobility (NTU, DT-SL, MDT and RoG) and the proposed measures (NT \--- group A; GCV-R and SV-R \--- group B; DT-R, TT-R and GCT-R \--- group C). For each of the 44 months between June 1, 2005 and January 31, 2009, we employ the spatiotemporal trajectories of callers during that month to estimate pairwise correlations between these 10 measures of mobility. We make use of the semiparametric Bayesian Gaussian copula estimation method of \cite{Hoff-2007} which produces estimates of the correlation matrix of the Gaussian copula via a Markov chain Monte Carlo (MCMC) algorithm. For each of the 44 months, we employed this MCMC algorithm to draw 10000 samples from the posterior distribution of the 10 dimensional correlation matrix of the mobility measures, after discarding 2500 samples as burn-in. Despite the large sample sizes available for each of the 44 months of data, estimating pairwise associations by sample correlation coefficients is not ideal due to the non-Gaussian nature of the univariate marginal distributions of the values of each mobility measure. These marginal distributions show large proportions of callers as having low mobility levels or not moving at all, and also smaller but significant proportions of callers having very high mobility levels. By employing a Gaussian copula, \cite{Hoff-2007} treats the univariate marginal distributions of measures of mobility as nuisance parameters, thus producing estimates of pairwise correlations with improved statistical properties.

\begin{figure}[H]
\begin{center}
\includegraphics[width=5.8in,angle=0]{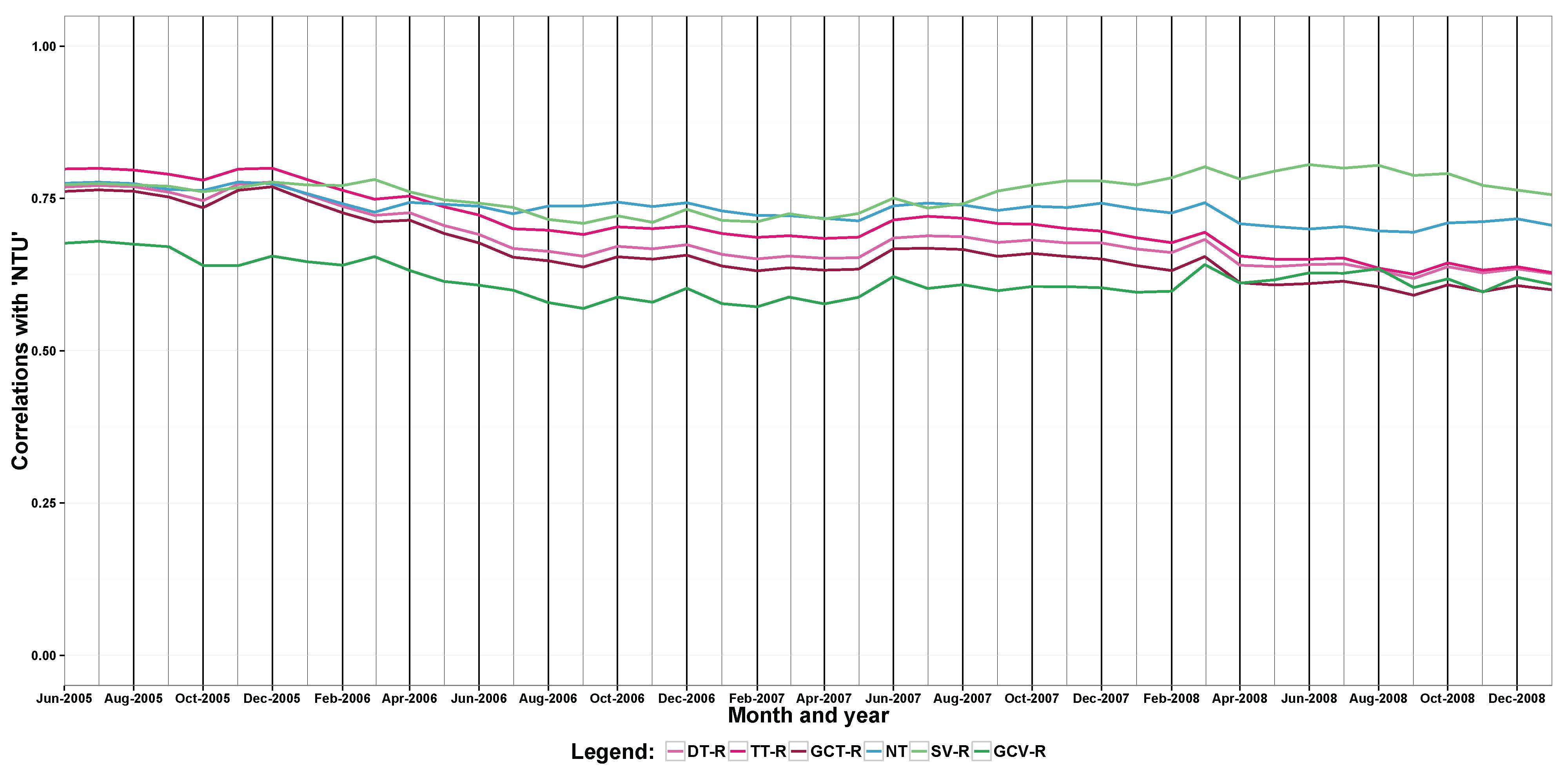}
\caption{Estimated correlations for each of the 44 months of Rwandan CDRs between the existing measure of mobility number of towers used (NTU) and the six measures from Groups A (\textcolor{blue}{blue}), B (shades of \textcolor{green}{green}) and C (shades of \textcolor{red}{red}).}
\label{fig:correlationsWithNTU}
\end{center}
\end{figure}

\begin{figure}[H]
\begin{center}
\includegraphics[width=5.8in,angle=0]{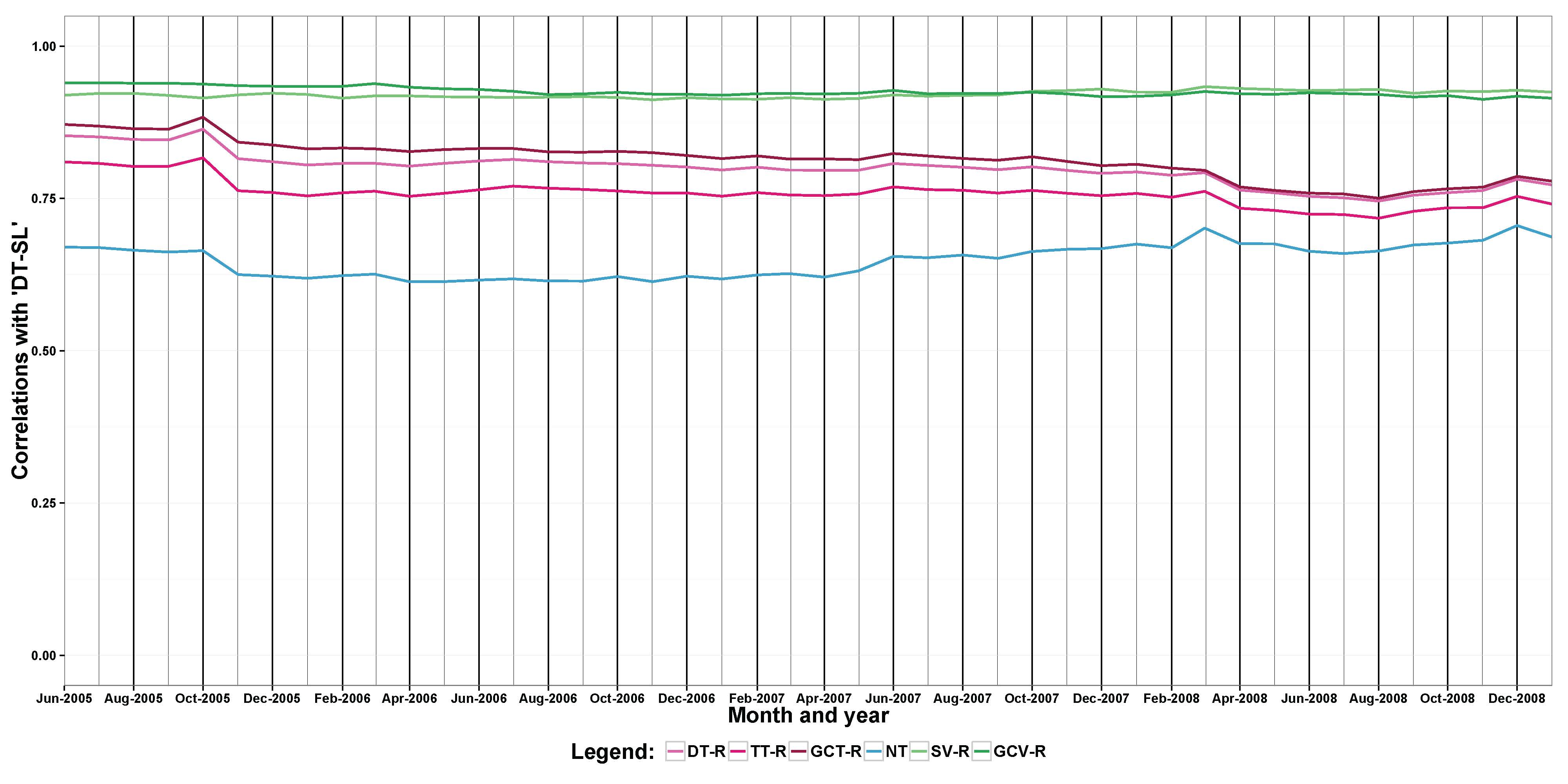}
\caption{Estimated correlations for each of the 44 months of Rwandan CDRs between the existing measure of mobility distance traveled (DT-SL) and the six measures from Groups A (\textcolor{blue}{blue}), B (shades of \textcolor{green}{green}) and C (shades of \textcolor{red}{red}).}
\label{fig:correlationsWithDT-SL}
\end{center}
\end{figure}

\begin{figure}[H]
\begin{center}
\includegraphics[width=5.8in,angle=0]{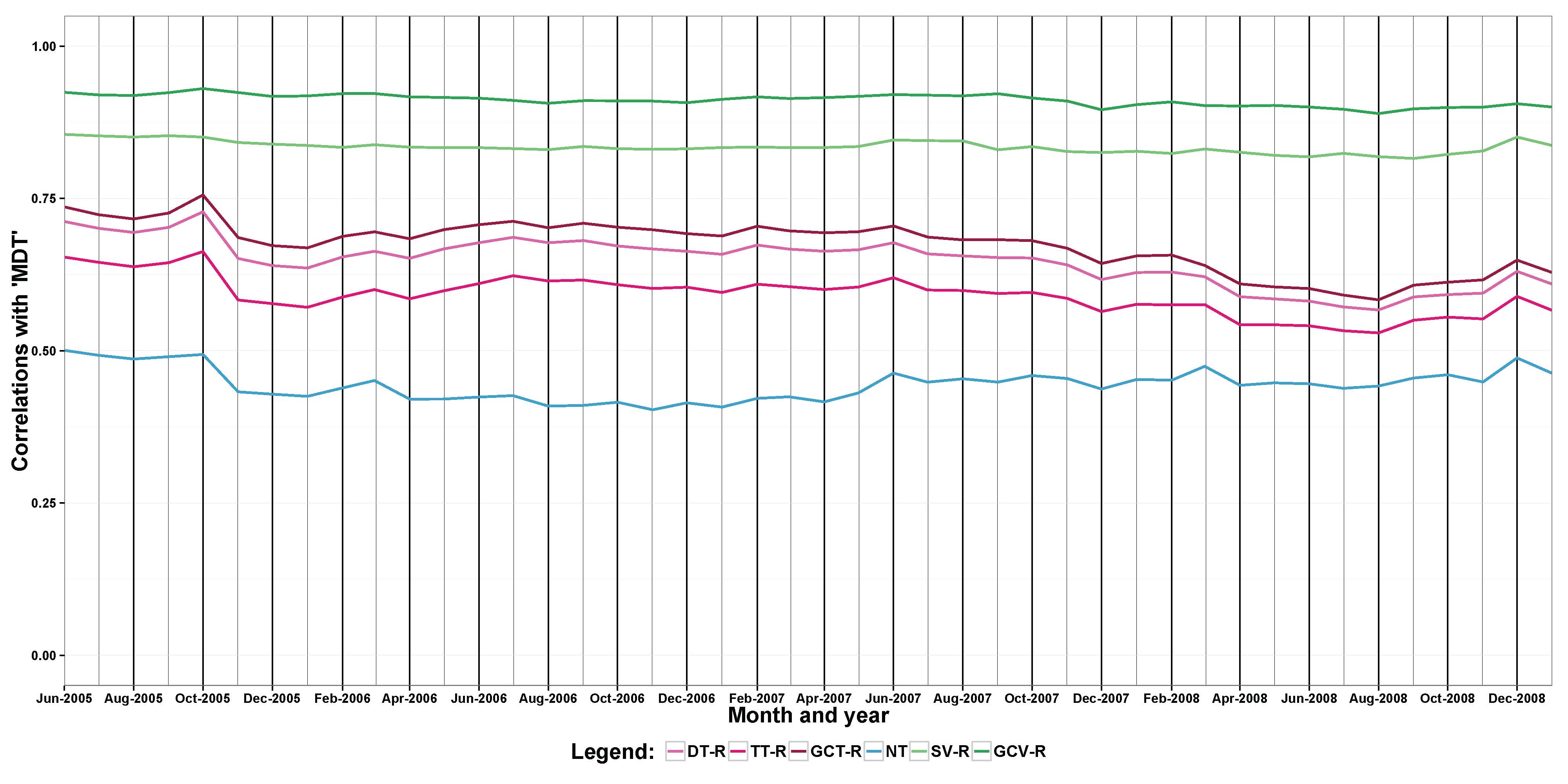}
\caption{Estimated correlations for each of the 44 months of Rwandan CDRs between the existing measure of mobility maximum distance traveled (MDT) and the six measures from Groups A (\textcolor{blue}{blue}), B (shades of \textcolor{green}{green}) and C (shades of \textcolor{red}{red}).}
\label{fig:correlationsWithMDT}
\end{center}
\end{figure}

\begin{figure}[H]
\begin{center}
\includegraphics[width=5.8in,angle=0]{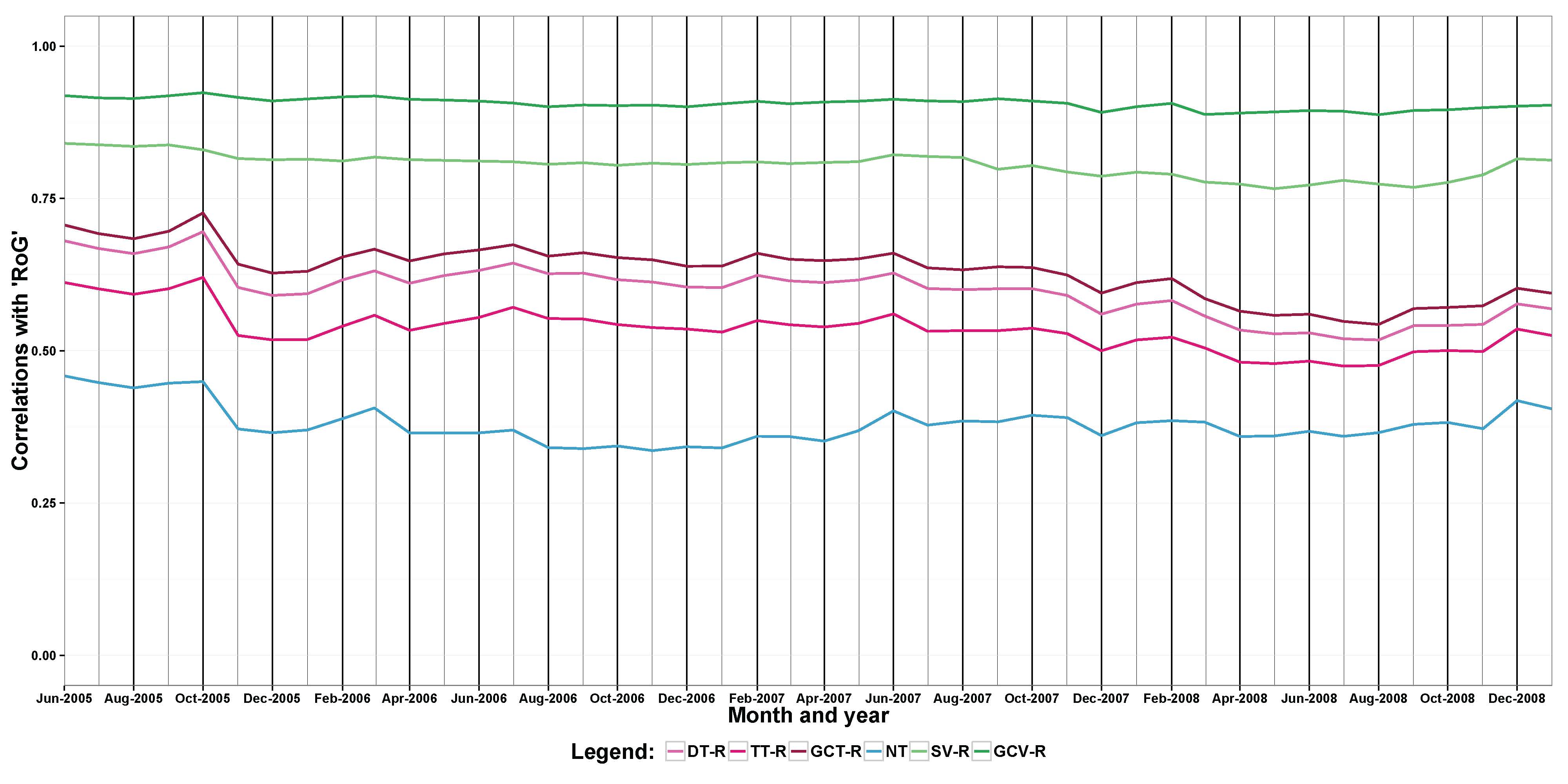}
\caption{Estimated correlations for each of the 44 months of Rwandan CDRs between the existing measure of mobility radius of gyration (RoG) and the six measures from Groups A (\textcolor{blue}{blue}), B (shades of \textcolor{green}{green}) and C (shades of \textcolor{red}{red}).}
\label{fig:correlationsWithRoG}
\end{center}
\end{figure}

\begin{figure}[H]
\begin{center}
\includegraphics[width=5.8in,angle=0]{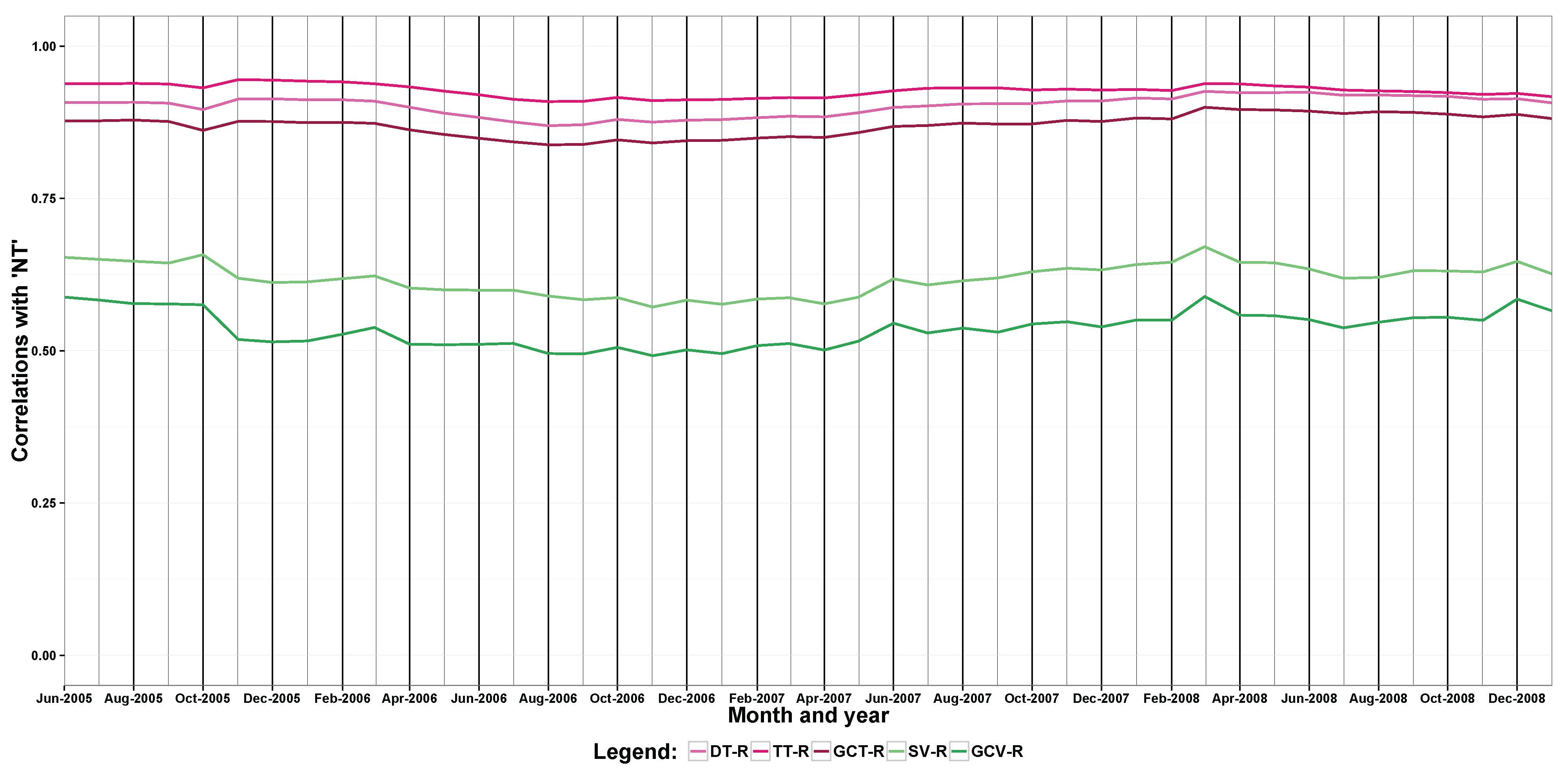}
\caption{Estimated correlations for each of the 44 months of Rwandan CDRs between the new measure of mobility number of trips (NT) which defines group A, and the five measures from Groups B (shades of \textcolor{green}{green}) and C (shades of \textcolor{red}{red}).}
\label{fig:correlationsWithNT}
\end{center}
\end{figure}

\begin{figure}[H]
\begin{center}
\includegraphics[width=5.8in,angle=0]{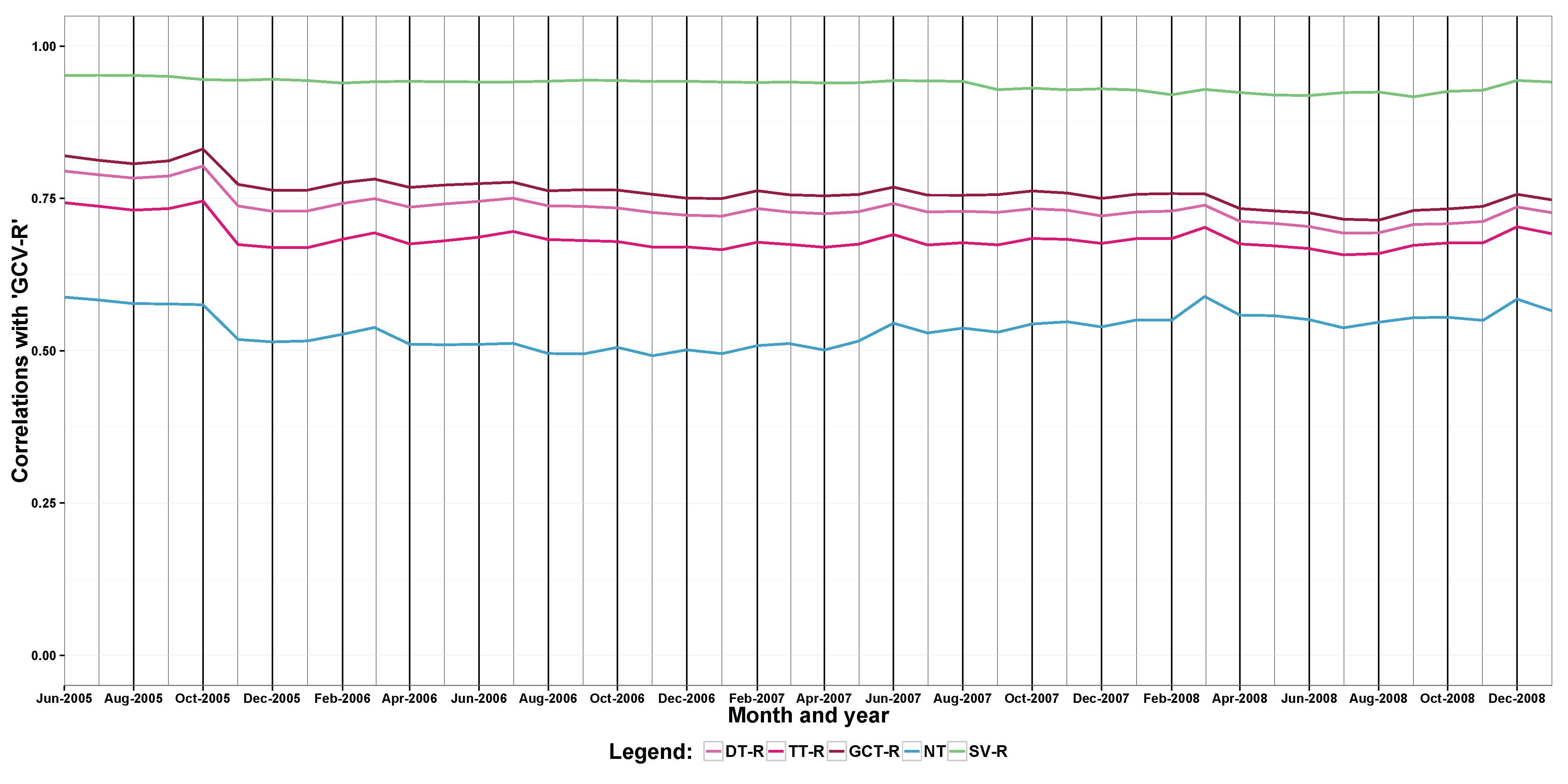}
\caption{Estimated correlations for each of the 44 months of Rwandan CDRs between the new measure of mobility grid cells visited (GCV-R) from group B, and the other measure of mobility from group B (SV-R, \textcolor{green}{green}), as well as the four measures from Groups A (\textcolor{blue}{blue}) and C (shades of \textcolor{red}{red}).}
\label{fig:correlationsWithGCV-R}
\end{center}
\end{figure}

\begin{figure}[H]
\begin{center}
\includegraphics[width=5.8in,angle=0]{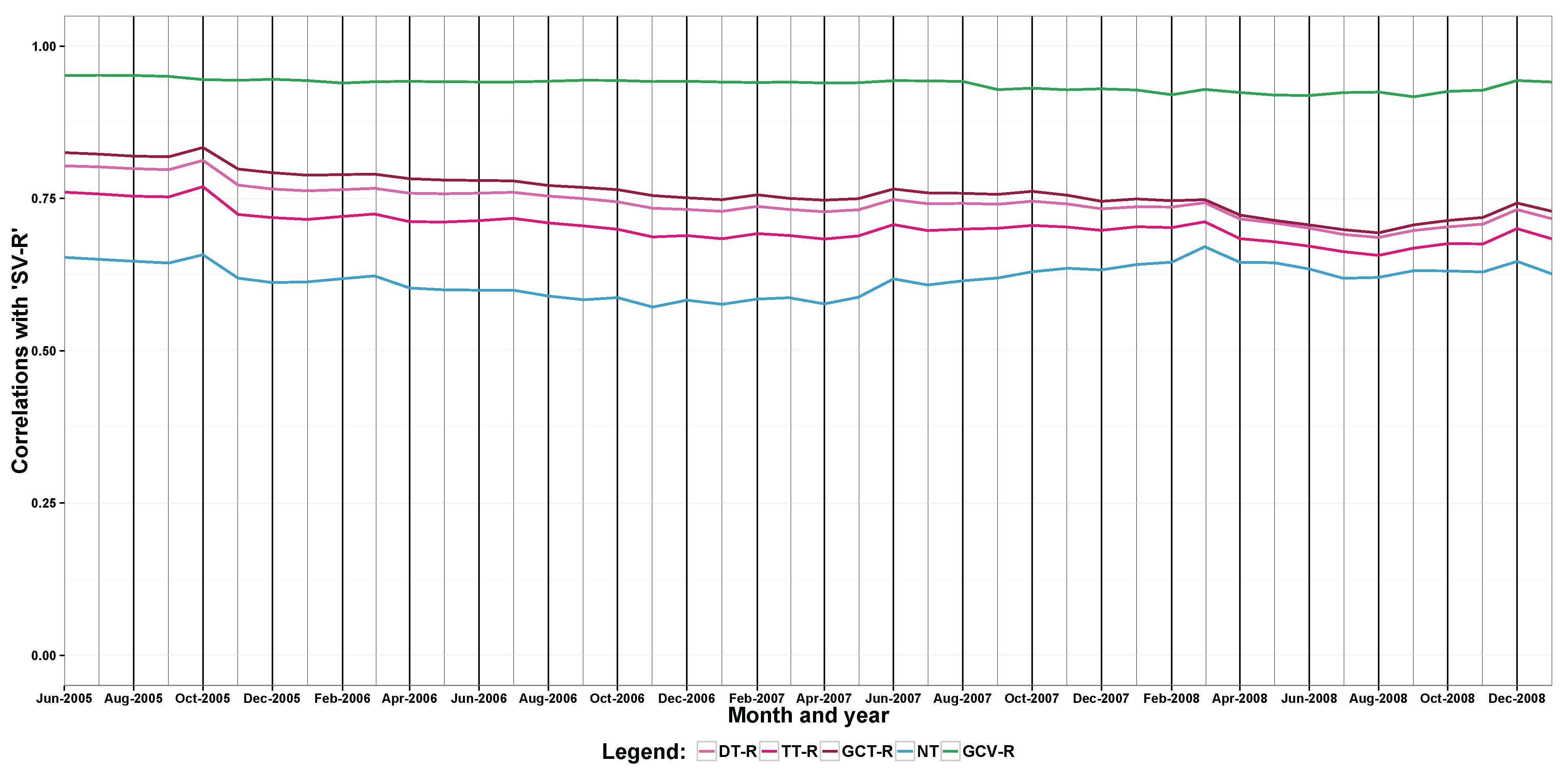}
\caption{Estimated correlations for each of the 44 months of Rwandan CDRs between the new measure of mobility sites visited (SV-R) from group B, and the other measure of mobility from group B (GCV-R, \textcolor{green}{green}), as well as the four measures from Groups A (\textcolor{blue}{blue}) and C (shades of \textcolor{red}{red}).}
\label{fig:correlationsWithSV-R}
\end{center}
\end{figure}

\begin{figure}[H]
\begin{center}
\includegraphics[width=5.8in,angle=0]{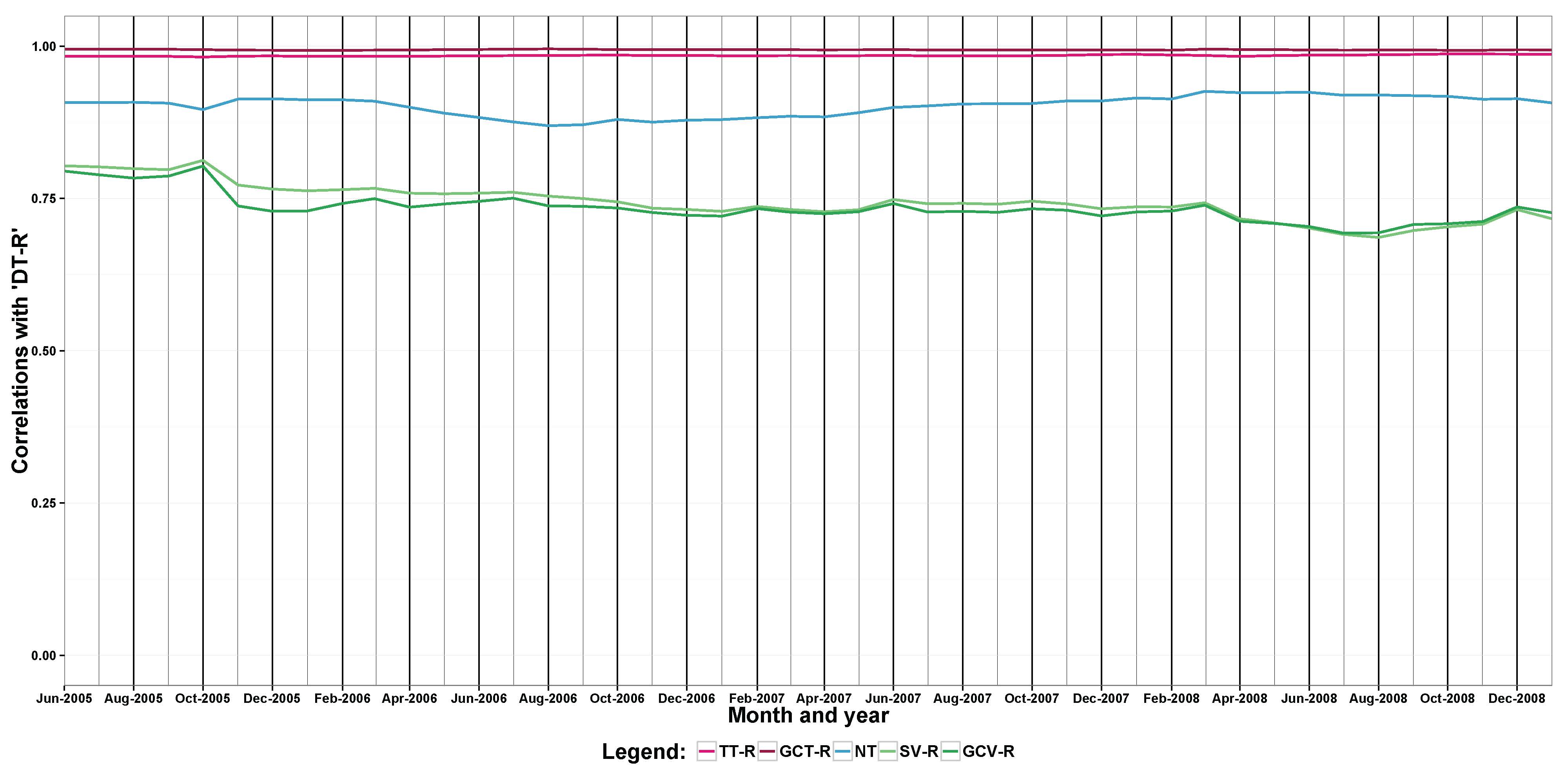}
\caption{Estimated correlations for each of the 44 months of Rwandan CDRs between the new measure of mobility distance traveled (DT-R) from group C, and the other measures of mobility from group C (TT-R and GCT-R, shades of \textcolor{red}{red}), as well as the three measures from Groups A (\textcolor{blue}{blue}) and B (shades of \textcolor{green}{green}).}
\label{fig:correlationsWithDT-R}
\end{center}
\end{figure}

\begin{figure}[H]
\begin{center}
\includegraphics[width=5.8in,angle=0]{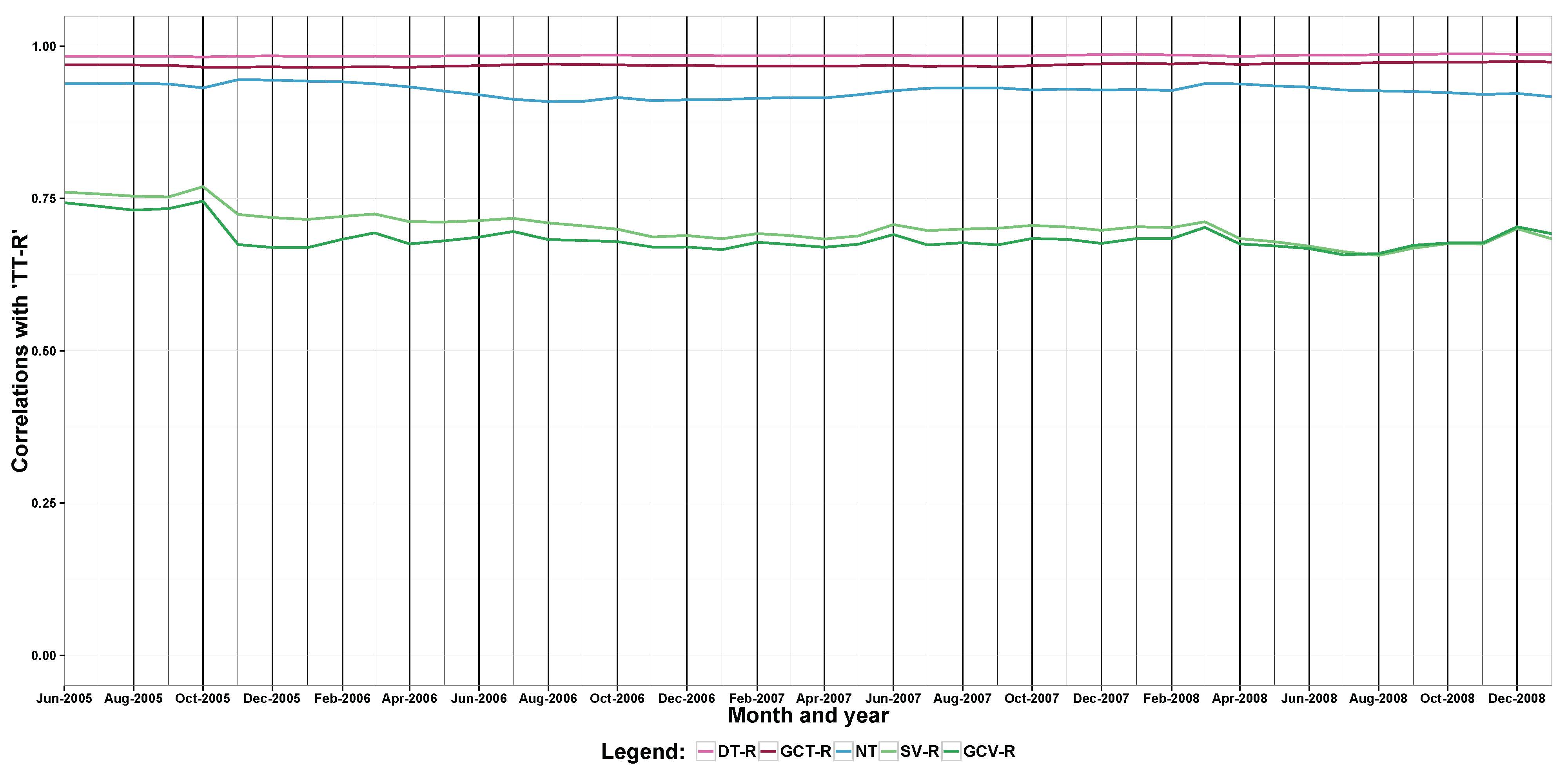}
\caption{Estimated correlations for each of the 44 months of Rwandan CDRs between the new measure of mobility time traveled (TT-R) from group C, and the other measures of mobility from group C (DT-R and GCT-R, shades of \textcolor{red}{red}), as well as the three measures from Groups A (\textcolor{blue}{blue}) and B (shades of \textcolor{green}{green}).}
\label{fig:correlationsWithTT-R}
\end{center}
\end{figure}

\begin{figure}[H]
\begin{center}
\includegraphics[width=5.8in,angle=0]{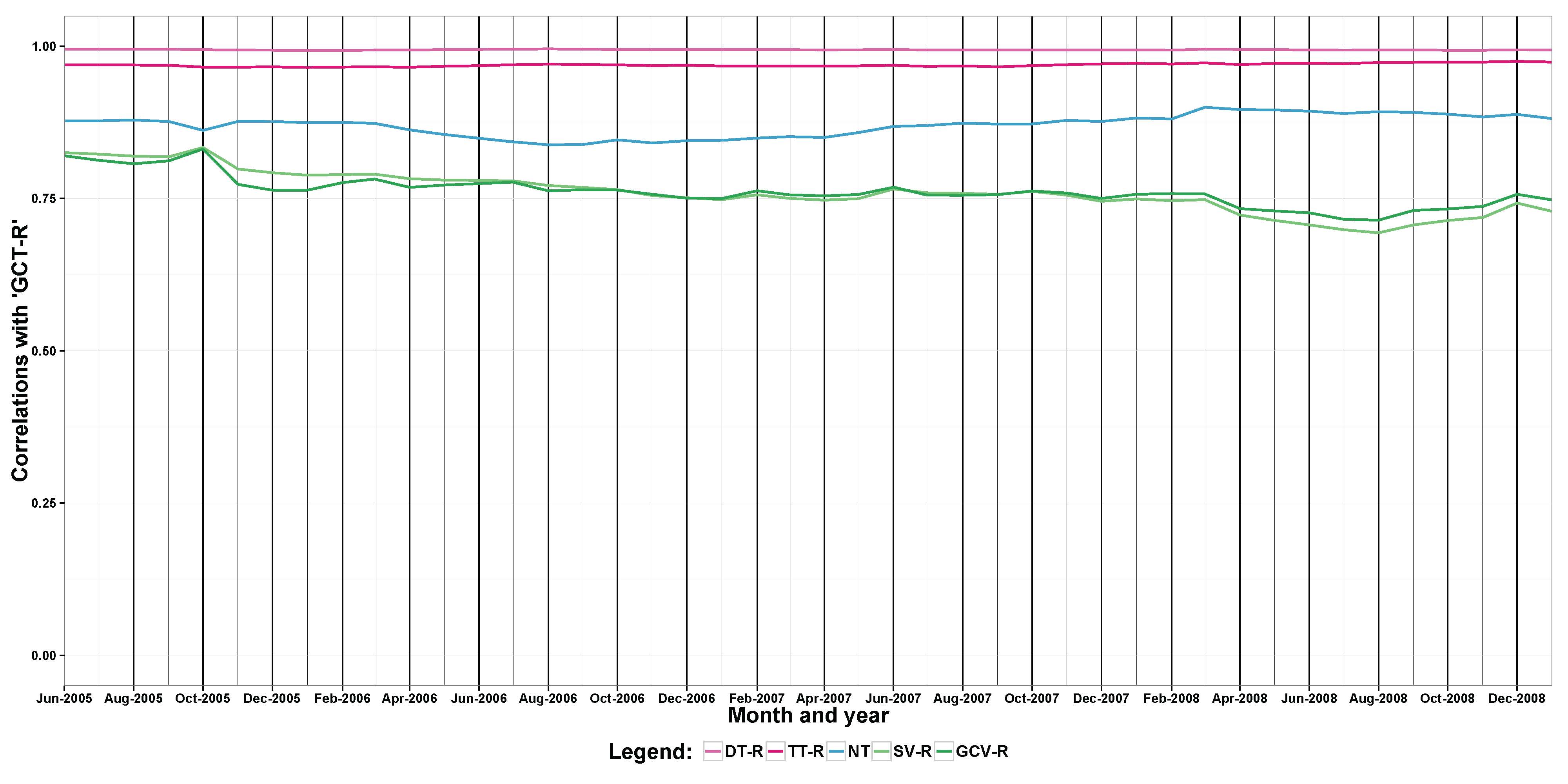}
\caption{Estimated correlations for each of the 44 months of Rwandan CDRs between the new measure of mobility grid cells traveled (GCT-R) from group C, and the other measures of mobility from group C (DT-R and TT-R, shades of \textcolor{red}{red}), as well as the three measures from Groups A (\textcolor{blue}{blue}) and B (shades of \textcolor{green}{green}).}
\label{fig:correlationsWithGCT-R}
\end{center}
\end{figure}

Figures \ref{fig:correlationsWithNTU}, \ref{fig:correlationsWithDT-SL}, \ref{fig:correlationsWithMDT} and \ref{fig:correlationsWithRoG} show the longitudinal pairwise correlations between each of the four existing measures of mobility and the six new measures from groups A, B and C. Figures \ref{fig:correlationsWithNT}, \ref{fig:correlationsWithGCV-R}, \ref{fig:correlationsWithSV-R}, \ref{fig:correlationsWithDT-R}, \ref{fig:correlationsWithTT-R} and \ref{fig:correlationsWithGCT-R} present the longitudinal pairwise associations between each of the six new measures of mobility and the other five measures from groups A, B and C. Since the three groups of measures were defined with respect to the two key dimensions of mobility (frequency of mobility and spatial range), it is important to observe the relationships between each measure with the other measures in the same group (only for groups B and C which contain more than one measure), and also with the measures in the remaining two groups. To this end, we used shades of the same color (\textcolor{green}{green} for group B and \textcolor{red}{red} for group C) to make it easier to identify correlations with measures in the same group. Correlations with the NT measure from group A are colored in \textcolor{blue}{blue}. The coloring of groups A, B and C is consistent in all the plots in this section (Figures \ref{fig:correlationsWithNTU}\---\ref{fig:correlationsWithGCT-R}).

Figure \ref{fig:correlationsWithNTU} shows that the existing measure of mobility number of towers used (NTU) has comparable correlations with the six new measures. The range of these correlations seem to be larger in 2008 compared to 2005, possibly due to the spatial range and the density of towers being significantly larger in 2008. Recall that tower density influences NTU but not our new measures. Notably, NTU seems to have the strongest longitudinal associations with the SV-R measure from group B which captures spatial range but does not capture the frequency of mobility. This is not surprising, since NTU also does not capture the frequency of mobility, and the manner in which it captures spatial range is confounded by the varying density of cellular towers. Nevertheless, NTU has the weakest association with the GCV-R measure, also from group B. As GCV-R counts grid cells on quickest road routes and SV-R counts only a subset of these grid cells which are also sites, we can see why NTU would be more similar to SV-R: NTU is based only from places in which calls were made. That said, NTU's correlations with the six new measures are consistent with NTU not being classified in any of the groups A, B and C.

Figures \ref{fig:correlationsWithDT-SL}, \ref{fig:correlationsWithMDT} and \ref{fig:correlationsWithRoG} show that the existing measures of mobility distance traveled (DT-SL), maximum distance traveled (MDT), and radius of gyration (RoG) share a common pattern of longitudinal associations with the six new measures of mobility: their strongest correlations are with the measures from group B, their second strongest associations are with the measures from group C, and their weakest associations are with the measure from group A. This is precisely as expected in the case of MDT and RoG since both measures capture information related to spatial range, but do not capture any aspect of the frequency of mobility which is key for the measures in groups A and C. However, this pattern of associations is somewhat surprising for DT-SL. We classified this measure in group C because it seemed to capture frequency of mobility and spatial range, and subsequently we expected to see the largest correlations with the new measures of mobility from group C, especially with distance traveled (DT-R). But there are key differences between DT-SL and DT-R: (i) DT-SL involves movement between cellular towers, while DT-R involves movement between sites; and (ii) DT-SL is the sum of straight line distances, while DT-R is the sum of distances via road travel. The lower longitudinal associations we observe between DT-SL and DT-R emphasize the key differences in the way these two measures are constructed.

Figure \ref{fig:correlationsWithNT} shows that the new measure of mobility number of trips (NT) which defines group A has the strongest longitudinal associations with the measures from group C, and weaker associations with the measures from group B.  This is consistent with our intuition about these measures: NT captures the frequency of mobility as do measures from group C. But NT does not capture spatial range of mobility which explains its lower correlations with the measures from group B.

Figures \ref{fig:correlationsWithGCV-R} and \ref{fig:correlationsWithSV-R} show that the two new measures of mobility from group B have strong longitudinal correlations with each other. Their second strongest longitudinal correlations are with the three measures from group C. This is consistent with our intuition since the five measures from group B and C capture spatial range. Their weakest longitudinal associations are estimated to be with the measure of mobility number of trips (NT) from group A. This is logical since NT captures only the frequency of mobility, while the two measures from group B capture only spatial range.

Figures \ref{fig:correlationsWithDT-R}, \ref{fig:correlationsWithTT-R} and \ref{fig:correlationsWithGCT-R} show that the three new measures of mobility from group C have strong longitudinal associations with each other. Their second strongest longitudinal associations are with the measure of mobility number of trips (NT) from group A. Their weakest longitudinal associations are estimated to be with the two measures from group B. The measures from group B capture spatial range as do the measures from group C, but they do not capture the frequency of mobility. Here it is interesting to see that frequency of mobility is one of the two key dimensions of mobility which leads to larger correlations between the measures in group C (which capture both dimensions) and the measures in groups A and B (which capture only one of these dimensions). 

\subsection*{SI6: Categories of callers defined by their mobility with respect to the measures in Groups A, B and C}

We identify categories of callers with respect to the six measures of mobility we propose in the main text.  A monthly spatiotemporal trajectory of a caller was classified as having low or high mobility with respect to a measure if the corresponding value of trajectory's measure is above or below the median of observed values during that month. For each month, this leads to $2^6=64$ segments. The segments which comprise more than $0.05\%$ of the callers in a given month are shown in Tables \ref{tab:dec2005}, \ref{tab:dec2006}, \ref{tab:dec2007} and \ref{tab:dec2008}. These tables are obtained by cross-classifying spatiotemporal trajectories observed in December 2005, December 2006, December 2007, and December 2008, respectively.

\begin{table}[H]
\caption{\label{tab:dec2005}Proportions of callers in categories defined by the six new measures of mobility from Groups A, B and C in December 2005. The measures are evaluated based on the spatiotemporal trajectories of each of the 238,572 persons that used a major cellular phone services provider's Rwandan network  during that month. From the $2^6=64$ possible categories, only the $37$ categories that contain at least $0.05\%$ of the callers from that month are shown.}
\begin{center} {\small\renewcommand{\arraystretch}{0.5}
\begin{tabular}{c|cc|ccc|rl}
{\bf Group A}& \multicolumn{2}{c|}{\bf Group B} & \multicolumn{3}{c|}{\bf Group C} & \multicolumn{2}{c}{\bf Percentage}\\ 
 NT &  GCV-R & SV-R & DT-R & TT-R & GCV-R & \\ \hline
\cellcolor{mattred!80} Low & \cellcolor{mattred!80} Low & \cellcolor{mattred!80} Low & \cellcolor{mattred!80} Low & \cellcolor{mattred!80} Low & \cellcolor{mattred!80} Low & 32.06& \mybar{0.3206}\\
\cellcolor{mattgreen!80} High & \cellcolor{mattgreen!80} High & \cellcolor{mattgreen!80} High & \cellcolor{mattgreen!80} High & \cellcolor{mattgreen!80} High & \cellcolor{mattgreen!80} High & 27.87& \mybar{0.2787}\\
\cellcolor{mattgreen!80} High & \cellcolor{mattred!80} Low & \cellcolor{mattred!80} Low & \cellcolor{mattgreen!80} High & \cellcolor{mattgreen!80} High & \cellcolor{mattgreen!80} High & 9.21& \mybar{0.0921}\\
\cellcolor{mattred!80} Low & \cellcolor{mattgreen!80} High & \cellcolor{mattgreen!80} High & \cellcolor{mattred!80} Low & \cellcolor{mattred!80} Low & \cellcolor{mattred!80} Low & 5.08& \mybar{0.0508}\\
\cellcolor{mattred!80} Low & \cellcolor{mattgreen!80} High & \cellcolor{mattred!80} Low & \cellcolor{mattred!80} Low & \cellcolor{mattred!80} Low & \cellcolor{mattred!80} Low & 4.59& \mybar{0.0459}\\
\cellcolor{mattred!80} Low & \cellcolor{mattgreen!80} High & \cellcolor{mattgreen!80} High & \cellcolor{mattgreen!80} High & \cellcolor{mattgreen!80} High & \cellcolor{mattgreen!80} High & 4.46& \mybar{0.0446}\\
\cellcolor{mattgreen!80} High & \cellcolor{mattred!80} Low & \cellcolor{mattred!80} Low & \cellcolor{mattred!80} Low & \cellcolor{mattred!80} Low & \cellcolor{mattred!80} Low & 3.59& \mybar{0.0359}\\
\cellcolor{mattgreen!80} High & \cellcolor{mattgreen!80} High & \cellcolor{mattred!80} Low & \cellcolor{mattgreen!80} High & \cellcolor{mattgreen!80} High & \cellcolor{mattgreen!80} High & 2.14& \mybar{0.0214}\\
\cellcolor{mattgreen!80} High & \cellcolor{mattred!80} Low & \cellcolor{mattred!80} Low & \cellcolor{mattred!80} Low & \cellcolor{mattgreen!80} High & \cellcolor{mattred!80} Low & 1.94& \mybar{0.0194}\\
\cellcolor{mattred!80} Low & \cellcolor{mattgreen!80} High & \cellcolor{mattgreen!80} High & \cellcolor{mattgreen!80} High & \cellcolor{mattred!80} Low & \cellcolor{mattgreen!80} High & 1.87& \mybar{0.0187}\\
\cellcolor{mattgreen!80} High & \cellcolor{mattred!80} Low & \cellcolor{mattred!80} Low & \cellcolor{mattgreen!80} High & \cellcolor{mattgreen!80} High & \cellcolor{mattred!80} Low & 1.31& \mybar{0.0131}\\
\cellcolor{mattgreen!80} High & \cellcolor{mattred!80} Low & \cellcolor{mattgreen!80} High & \cellcolor{mattgreen!80} High & \cellcolor{mattgreen!80} High & \cellcolor{mattgreen!80} High & 1.02& \mybar{0.0102}\\
\cellcolor{mattred!80} Low & \cellcolor{mattgreen!80} High & \cellcolor{mattgreen!80} High & \cellcolor{mattred!80} Low & \cellcolor{mattred!80} Low & \cellcolor{mattgreen!80} High & 0.94& \mybar{0.0094}\\
\cellcolor{mattred!80} Low & \cellcolor{mattgreen!80} High & \cellcolor{mattred!80} Low & \cellcolor{mattgreen!80} High & \cellcolor{mattgreen!80} High & \cellcolor{mattgreen!80} High & 0.74& \mybar{0.0074}\\
\cellcolor{mattred!80} Low & \cellcolor{mattgreen!80} High & \cellcolor{mattred!80} Low & \cellcolor{mattgreen!80} High & \cellcolor{mattred!80} Low & \cellcolor{mattgreen!80} High & 0.30& \mybar{0.003}\\
\cellcolor{mattgreen!80} High & \cellcolor{mattgreen!80} High & \cellcolor{mattgreen!80} High & \cellcolor{mattgreen!80} High & \cellcolor{mattred!80} Low & \cellcolor{mattgreen!80} High & 0.29& \mybar{0.0029}\\
\cellcolor{mattgreen!80} High & \cellcolor{mattgreen!80} High & \cellcolor{mattgreen!80} High & \cellcolor{mattred!80} Low & \cellcolor{mattred!80} Low & \cellcolor{mattred!80} Low & 0.25& \mybar{0.0025}\\
\cellcolor{mattred!80} Low & \cellcolor{mattgreen!80} High & \cellcolor{mattred!80} Low & \cellcolor{mattred!80} Low & \cellcolor{mattred!80} Low & \cellcolor{mattgreen!80} High & 0.20& \mybar{0.002}\\
\cellcolor{mattgreen!80} High & \cellcolor{mattgreen!80} High & \cellcolor{mattred!80} Low & \cellcolor{mattred!80} Low & \cellcolor{mattred!80} Low & \cellcolor{mattred!80} Low & 0.16& \mybar{0.0016}\\
\cellcolor{mattred!80} Low & \cellcolor{mattgreen!80} High & \cellcolor{mattred!80} Low & \cellcolor{mattred!80} Low & \cellcolor{mattgreen!80} High & \cellcolor{mattred!80} Low & 0.15& \mybar{0.0015}\\
\cellcolor{mattred!80} Low & \cellcolor{mattred!80} Low & \cellcolor{mattred!80} Low & \cellcolor{mattred!80} Low & \cellcolor{mattgreen!80} High & \cellcolor{mattred!80} Low & 0.14& \mybar{0.0014}\\
\cellcolor{mattred!80} Low & \cellcolor{mattred!80} Low & \cellcolor{mattgreen!80} High & \cellcolor{mattred!80} Low & \cellcolor{mattred!80} Low & \cellcolor{mattred!80} Low & 0.12& \mybar{0.0012}\\
\cellcolor{mattred!80} Low & \cellcolor{mattred!80} Low & \cellcolor{mattred!80} Low & \cellcolor{mattgreen!80} High & \cellcolor{mattgreen!80} High & \cellcolor{mattgreen!80} High & 0.12& \mybar{0.0012}\\
\cellcolor{mattred!80} Low & \cellcolor{mattgreen!80} High & \cellcolor{mattred!80} Low & \cellcolor{mattgreen!80} High & \cellcolor{mattgreen!80} High & \cellcolor{mattred!80} Low & 0.11& \mybar{0.0011}\\
\cellcolor{mattgreen!80} High & \cellcolor{mattgreen!80} High & \cellcolor{mattred!80} Low & \cellcolor{mattred!80} Low & \cellcolor{mattgreen!80} High & \cellcolor{mattred!80} Low & 0.10& \mybar{0.001}\\
\cellcolor{mattgreen!80} High & \cellcolor{mattgreen!80} High & \cellcolor{mattgreen!80} High & \cellcolor{mattred!80} Low & \cellcolor{mattgreen!80} High & \cellcolor{mattred!80} Low & 0.10& \mybar{0.001}\\
\cellcolor{mattgreen!80} High & \cellcolor{mattgreen!80} High & \cellcolor{mattred!80} Low & \cellcolor{mattgreen!80} High & \cellcolor{mattgreen!80} High & \cellcolor{mattred!80} Low & 0.09& \mybar{0.0009}\\
\cellcolor{mattred!80} Low & \cellcolor{mattgreen!80} High & \cellcolor{mattgreen!80} High & \cellcolor{mattred!80} Low & \cellcolor{mattgreen!80} High & \cellcolor{mattred!80} Low & 0.08& \mybar{0.0008}\\
\cellcolor{mattgreen!80} High & \cellcolor{mattred!80} Low & \cellcolor{mattred!80} Low & \cellcolor{mattgreen!80} High & \cellcolor{mattred!80} Low & \cellcolor{mattgreen!80} High & 0.08& \mybar{0.0008}\\
\cellcolor{mattred!80} Low & \cellcolor{mattred!80} Low & \cellcolor{mattred!80} Low & \cellcolor{mattred!80} Low & \cellcolor{mattred!80} Low & \cellcolor{mattgreen!80} High & 0.08& \mybar{0.0008}\\
\cellcolor{mattgreen!80} High & \cellcolor{mattgreen!80} High & \cellcolor{mattgreen!80} High & \cellcolor{mattgreen!80} High & \cellcolor{mattgreen!80} High & \cellcolor{mattred!80} Low & 0.08& \mybar{0.0008}\\
\cellcolor{mattgreen!80} High & \cellcolor{mattred!80} Low & \cellcolor{mattgreen!80} High & \cellcolor{mattred!80} Low & \cellcolor{mattgreen!80} High & \cellcolor{mattred!80} Low & 0.07& \mybar{0.0007}\\
\cellcolor{mattgreen!80} High & \cellcolor{mattred!80} Low & \cellcolor{mattred!80} Low & \cellcolor{mattred!80} Low & \cellcolor{mattred!80} Low & \cellcolor{mattgreen!80} High & 0.07& \mybar{0.0007}\\
\cellcolor{mattgreen!80} High & \cellcolor{mattgreen!80} High & \cellcolor{mattred!80} Low & \cellcolor{mattgreen!80} High & \cellcolor{mattred!80} Low & \cellcolor{mattgreen!80} High & 0.06& \mybar{0.0006}\\
\cellcolor{mattred!80} Low & \cellcolor{mattred!80} Low & \cellcolor{mattred!80} Low & \cellcolor{mattgreen!80} High & \cellcolor{mattred!80} Low & \cellcolor{mattgreen!80} High & 0.05& \mybar{0.0005}\\
\cellcolor{mattgreen!80} High & \cellcolor{mattred!80} Low & \cellcolor{mattgreen!80} High & \cellcolor{mattred!80} Low & \cellcolor{mattred!80} Low & \cellcolor{mattred!80} Low & 0.05& \mybar{0.0005}\\
\cellcolor{mattgreen!80} High & \cellcolor{mattred!80} Low & \cellcolor{mattgreen!80} High & \cellcolor{mattgreen!80} High & \cellcolor{mattgreen!80} High & \cellcolor{mattred!80} Low & 0.05& \mybar{0.0005}\\
\end{tabular} }
\end{center}
\end{table}

\begin{table}[H]
\caption{\label{tab:dec2006}Proportions of callers in categories defined by the six new measures of mobility from Groups A, B and C in December 2006. The measures are evaluated based on the spatiotemporal trajectories of each of the 310,877 persons that used a major cellular phone services provider's Rwandan network  during that month. From the $2^6=64$ possible categories, only the $39$ categories that contain at least $0.05\%$ of the callers from that month are shown.}
\begin{center} {\small\renewcommand{\arraystretch}{0.5}
\begin{tabular}{c|cc|ccc|rl}
{\bf Group A}& \multicolumn{2}{c|}{\bf Group B} & \multicolumn{3}{c|}{\bf Group C} & \multicolumn{2}{c}{\bf Percentage}\\ 
 NT &  GCV-R & SV-R & DT-R & TT-R & GCV-R & \\ \hline
\cellcolor{mattred!80} Low & \cellcolor{mattred!80} Low & \cellcolor{mattred!80} Low & \cellcolor{mattred!80} Low & \cellcolor{mattred!80} Low & \cellcolor{mattred!80} Low & 31.11& \mybar{0.3111}\\
\cellcolor{mattgreen!80} High & \cellcolor{mattgreen!80} High & \cellcolor{mattgreen!80} High & \cellcolor{mattgreen!80} High & \cellcolor{mattgreen!80} High & \cellcolor{mattgreen!80} High & 29.68& \mybar{0.2968}\\
\cellcolor{mattgreen!80} High & \cellcolor{mattred!80} Low & \cellcolor{mattred!80} Low & \cellcolor{mattgreen!80} High & \cellcolor{mattgreen!80} High & \cellcolor{mattgreen!80} High & 8.75& \mybar{0.0875}\\
\cellcolor{mattred!80} Low & \cellcolor{mattgreen!80} High & \cellcolor{mattgreen!80} High & \cellcolor{mattred!80} Low & \cellcolor{mattred!80} Low & \cellcolor{mattred!80} Low & 7.97& \mybar{0.0797}\\
\cellcolor{mattred!80} Low & \cellcolor{mattgreen!80} High & \cellcolor{mattgreen!80} High & \cellcolor{mattgreen!80} High & \cellcolor{mattgreen!80} High & \cellcolor{mattgreen!80} High & 3.78& \mybar{0.0378}\\
\cellcolor{mattgreen!80} High & \cellcolor{mattred!80} Low & \cellcolor{mattred!80} Low & \cellcolor{mattred!80} Low & \cellcolor{mattred!80} Low & \cellcolor{mattred!80} Low & 3.07& \mybar{0.0307}\\
\cellcolor{mattred!80} Low & \cellcolor{mattgreen!80} High & \cellcolor{mattred!80} Low & \cellcolor{mattred!80} Low & \cellcolor{mattred!80} Low & \cellcolor{mattred!80} Low & 2.08& \mybar{0.0208}\\
\cellcolor{mattgreen!80} High & \cellcolor{mattred!80} Low & \cellcolor{mattred!80} Low & \cellcolor{mattred!80} Low & \cellcolor{mattgreen!80} High & \cellcolor{mattred!80} Low & 1.98& \mybar{0.0198}\\
\cellcolor{mattred!80} Low & \cellcolor{mattgreen!80} High & \cellcolor{mattgreen!80} High & \cellcolor{mattgreen!80} High & \cellcolor{mattred!80} Low & \cellcolor{mattgreen!80} High & 1.55& \mybar{0.0155}\\
\cellcolor{mattgreen!80} High & \cellcolor{mattred!80} Low & \cellcolor{mattgreen!80} High & \cellcolor{mattgreen!80} High & \cellcolor{mattgreen!80} High & \cellcolor{mattgreen!80} High & 1.40& \mybar{0.014}\\
\cellcolor{mattgreen!80} High & \cellcolor{mattred!80} Low & \cellcolor{mattred!80} Low & \cellcolor{mattgreen!80} High & \cellcolor{mattgreen!80} High & \cellcolor{mattred!80} Low & 1.23& \mybar{0.0123}\\
\cellcolor{mattgreen!80} High & \cellcolor{mattgreen!80} High & \cellcolor{mattred!80} Low & \cellcolor{mattgreen!80} High & \cellcolor{mattgreen!80} High & \cellcolor{mattgreen!80} High & 1.06& \mybar{0.0106}\\
\cellcolor{mattred!80} Low & \cellcolor{mattgreen!80} High & \cellcolor{mattgreen!80} High & \cellcolor{mattred!80} Low & \cellcolor{mattred!80} Low & \cellcolor{mattgreen!80} High & 0.88& \mybar{0.0088}\\
\cellcolor{mattgreen!80} High & \cellcolor{mattgreen!80} High & \cellcolor{mattgreen!80} High & \cellcolor{mattred!80} Low & \cellcolor{mattred!80} Low & \cellcolor{mattred!80} Low & 0.68& \mybar{0.0068}\\
\cellcolor{mattred!80} Low & \cellcolor{mattred!80} Low & \cellcolor{mattgreen!80} High & \cellcolor{mattred!80} Low & \cellcolor{mattred!80} Low & \cellcolor{mattred!80} Low & 0.59& \mybar{0.0059}\\
\cellcolor{mattgreen!80} High & \cellcolor{mattgreen!80} High & \cellcolor{mattgreen!80} High & \cellcolor{mattgreen!80} High & \cellcolor{mattred!80} Low & \cellcolor{mattgreen!80} High & 0.51& \mybar{0.0051}\\
\cellcolor{mattred!80} Low & \cellcolor{mattgreen!80} High & \cellcolor{mattred!80} Low & \cellcolor{mattgreen!80} High & \cellcolor{mattgreen!80} High & \cellcolor{mattgreen!80} High & 0.43& \mybar{0.0043}\\
\cellcolor{mattred!80} Low & \cellcolor{mattred!80} Low & \cellcolor{mattred!80} Low & \cellcolor{mattgreen!80} High & \cellcolor{mattgreen!80} High & \cellcolor{mattgreen!80} High & 0.40& \mybar{0.004}\\
\cellcolor{mattred!80} Low & \cellcolor{mattred!80} Low & \cellcolor{mattred!80} Low & \cellcolor{mattgreen!80} High & \cellcolor{mattred!80} Low & \cellcolor{mattgreen!80} High & 0.24& \mybar{0.0024}\\
\cellcolor{mattgreen!80} High & \cellcolor{mattgreen!80} High & \cellcolor{mattgreen!80} High & \cellcolor{mattred!80} Low & \cellcolor{mattgreen!80} High & \cellcolor{mattred!80} Low & 0.23& \mybar{0.0023}\\
\cellcolor{mattred!80} Low & \cellcolor{mattred!80} Low & \cellcolor{mattred!80} Low & \cellcolor{mattred!80} Low & \cellcolor{mattred!80} Low & \cellcolor{mattgreen!80} High & 0.23& \mybar{0.0023}\\
\cellcolor{mattgreen!80} High & \cellcolor{mattgreen!80} High & \cellcolor{mattgreen!80} High & \cellcolor{mattgreen!80} High & \cellcolor{mattgreen!80} High & \cellcolor{mattred!80} Low & 0.23& \mybar{0.0023}\\
\cellcolor{mattgreen!80} High & \cellcolor{mattred!80} Low & \cellcolor{mattred!80} Low & \cellcolor{mattgreen!80} High & \cellcolor{mattred!80} Low & \cellcolor{mattgreen!80} High & 0.21& \mybar{0.0021}\\
\cellcolor{mattred!80} Low & \cellcolor{mattred!80} Low & \cellcolor{mattred!80} Low & \cellcolor{mattred!80} Low & \cellcolor{mattgreen!80} High & \cellcolor{mattred!80} Low & 0.15& \mybar{0.0015}\\
\cellcolor{mattgreen!80} High & \cellcolor{mattgreen!80} High & \cellcolor{mattred!80} Low & \cellcolor{mattred!80} Low & \cellcolor{mattred!80} Low & \cellcolor{mattred!80} Low & 0.12& \mybar{0.0012}\\
\cellcolor{mattgreen!80} High & \cellcolor{mattred!80} Low & \cellcolor{mattgreen!80} High & \cellcolor{mattred!80} Low & \cellcolor{mattred!80} Low & \cellcolor{mattred!80} Low & 0.11& \mybar{0.0011}\\
\cellcolor{mattred!80} Low & \cellcolor{mattgreen!80} High & \cellcolor{mattred!80} Low & \cellcolor{mattgreen!80} High & \cellcolor{mattred!80} Low & \cellcolor{mattgreen!80} High & 0.11& \mybar{0.0011}\\
\cellcolor{mattred!80} Low & \cellcolor{mattgreen!80} High & \cellcolor{mattred!80} Low & \cellcolor{mattred!80} Low & \cellcolor{mattred!80} Low & \cellcolor{mattgreen!80} High & 0.11& \mybar{0.0011}\\
\cellcolor{mattgreen!80} High & \cellcolor{mattred!80} Low & \cellcolor{mattgreen!80} High & \cellcolor{mattred!80} Low & \cellcolor{mattgreen!80} High & \cellcolor{mattred!80} Low & 0.10 & \mybar{0.001}\\
\cellcolor{mattred!80} Low & \cellcolor{mattgreen!80} High & \cellcolor{mattgreen!80} High & \cellcolor{mattred!80} Low & \cellcolor{mattgreen!80} High & \cellcolor{mattred!80} Low & 0.09& \mybar{0.0009}\\
\cellcolor{mattgreen!80} High & \cellcolor{mattred!80} Low & \cellcolor{mattred!80} Low & \cellcolor{mattred!80} Low & \cellcolor{mattred!80} Low & \cellcolor{mattgreen!80} High & 0.08& \mybar{0.0008}\\
\cellcolor{mattgreen!80} High & \cellcolor{mattred!80} Low & \cellcolor{mattgreen!80} High & \cellcolor{mattgreen!80} High & \cellcolor{mattgreen!80} High & \cellcolor{mattred!80} Low & 0.08& \mybar{0.0008}\\
\cellcolor{mattgreen!80} High & \cellcolor{mattgreen!80} High & \cellcolor{mattgreen!80} High & \cellcolor{mattred!80} Low & \cellcolor{mattred!80} Low & \cellcolor{mattgreen!80} High & 0.07& \mybar{0.0007}\\
\cellcolor{mattred!80} Low & \cellcolor{mattgreen!80} High & \cellcolor{mattred!80} Low & \cellcolor{mattred!80} Low & \cellcolor{mattgreen!80} High & \cellcolor{mattred!80} Low & 0.06& \mybar{0.0006}\\
\cellcolor{mattgreen!80} High & \cellcolor{mattgreen!80} High & \cellcolor{mattgreen!80} High & \cellcolor{mattgreen!80} High & \cellcolor{mattred!80} Low & \cellcolor{mattred!80} Low & 0.06& \mybar{0.0006}\\
\cellcolor{mattred!80} Low & \cellcolor{mattgreen!80} High & \cellcolor{mattgreen!80} High & \cellcolor{mattred!80} Low & \cellcolor{mattgreen!80} High & \cellcolor{mattgreen!80} High & 0.05& \mybar{0.0005}\\
\cellcolor{mattgreen!80} High & \cellcolor{mattred!80} Low & \cellcolor{mattred!80} Low & \cellcolor{mattgreen!80} High & \cellcolor{mattred!80} Low & \cellcolor{mattred!80} Low & 0.05& \mybar{0.0005}\\
\cellcolor{mattred!80} Low & \cellcolor{mattred!80} Low & \cellcolor{mattred!80} Low & \cellcolor{mattred!80} Low & \cellcolor{mattgreen!80} High & \cellcolor{mattgreen!80} High & 0.05& \mybar{0.0005}\\
\cellcolor{mattgreen!80} High & \cellcolor{mattgreen!80} High & \cellcolor{mattgreen!80} High & \cellcolor{mattred!80} Low & \cellcolor{mattgreen!80} High & \cellcolor{mattgreen!80} High & 0.05& \mybar{0.0005}\\
\end{tabular} }
\end{center}
\end{table}

\begin{table}[H]
\caption{\label{tab:dec2007}Proportions of callers in categories defined by the six new measures of mobility from Groups A, B and C in December 2007. The measures are evaluated based on the spatiotemporal trajectories of each of the 552,041 persons that used a major cellular phone services provider's Rwandan network  during that month. From the $2^6=64$ possible categories, only the $39$ categories that contain at least $0.05\%$ of the callers from that month are shown.}
\begin{center} {\small \renewcommand{\arraystretch}{0.5}
\begin{tabular}{c|cc|ccc|rl}
{\bf Group A}& \multicolumn{2}{c|}{\bf Group B} & \multicolumn{3}{c|}{\bf Group C} & \multicolumn{2}{c}{\bf Percentage}\\ 
 NT &  GCV-R & SV-R & DT-R & TT-R & GCV-R & \\ \hline
\cellcolor{mattred!80} Low & \cellcolor{mattred!80} Low & \cellcolor{mattred!80} Low & \cellcolor{mattred!80} Low & \cellcolor{mattred!80} Low & \cellcolor{mattred!80} Low & 31.28& \mybar{0.3128}\\
\cellcolor{mattgreen!80} High & \cellcolor{mattgreen!80} High & \cellcolor{mattgreen!80} High & \cellcolor{mattgreen!80} High & \cellcolor{mattgreen!80} High & \cellcolor{mattgreen!80} High & 28.77& \mybar{0.2877}\\
\cellcolor{mattred!80} Low & \cellcolor{mattgreen!80} High & \cellcolor{mattgreen!80} High & \cellcolor{mattred!80} Low & \cellcolor{mattred!80} Low & \cellcolor{mattred!80} Low & 7.47& \mybar{0.0747}\\
\cellcolor{mattgreen!80} High & \cellcolor{mattred!80} Low & \cellcolor{mattred!80} Low & \cellcolor{mattgreen!80} High & \cellcolor{mattgreen!80} High & \cellcolor{mattgreen!80} High & 6.70& \mybar{0.067}\\
\cellcolor{mattgreen!80} High & \cellcolor{mattred!80} Low & \cellcolor{mattgreen!80} High & \cellcolor{mattgreen!80} High & \cellcolor{mattgreen!80} High & \cellcolor{mattgreen!80} High & 3.53& \mybar{0.0353}\\
\cellcolor{mattred!80} Low & \cellcolor{mattgreen!80} High & \cellcolor{mattgreen!80} High & \cellcolor{mattgreen!80} High & \cellcolor{mattgreen!80} High & \cellcolor{mattgreen!80} High & 3.24& \mybar{0.0324}\\
\cellcolor{mattgreen!80} High & \cellcolor{mattred!80} Low & \cellcolor{mattred!80} Low & \cellcolor{mattred!80} Low & \cellcolor{mattred!80} Low & \cellcolor{mattred!80} Low & 2.86& \mybar{0.0286}\\
\cellcolor{mattgreen!80} High & \cellcolor{mattgreen!80} High & \cellcolor{mattred!80} Low & \cellcolor{mattgreen!80} High & \cellcolor{mattgreen!80} High & \cellcolor{mattgreen!80} High & 2.14& \mybar{0.0214}\\
\cellcolor{mattred!80} Low & \cellcolor{mattgreen!80} High & \cellcolor{mattred!80} Low & \cellcolor{mattred!80} Low & \cellcolor{mattred!80} Low & \cellcolor{mattred!80} Low & 2.10& \mybar{0.021}\\
\cellcolor{mattgreen!80} High & \cellcolor{mattred!80} Low & \cellcolor{mattred!80} Low & \cellcolor{mattred!80} Low & \cellcolor{mattgreen!80} High & \cellcolor{mattred!80} Low & 1.38& \mybar{0.0138}\\
\cellcolor{mattred!80} Low & \cellcolor{mattgreen!80} High & \cellcolor{mattgreen!80} High & \cellcolor{mattgreen!80} High & \cellcolor{mattred!80} Low & \cellcolor{mattgreen!80} High & 1.31& \mybar{0.0131}\\
\cellcolor{mattred!80} Low & \cellcolor{mattred!80} Low & \cellcolor{mattgreen!80} High & \cellcolor{mattred!80} Low & \cellcolor{mattred!80} Low & \cellcolor{mattred!80} Low & 1.20& \mybar{0.012}\\
\cellcolor{mattgreen!80} High & \cellcolor{mattred!80} Low & \cellcolor{mattred!80} Low & \cellcolor{mattgreen!80} High & \cellcolor{mattgreen!80} High & \cellcolor{mattred!80} Low & 1.04& \mybar{0.0104}\\
\cellcolor{mattred!80} Low & \cellcolor{mattgreen!80} High & \cellcolor{mattgreen!80} High & \cellcolor{mattred!80} Low & \cellcolor{mattred!80} Low & \cellcolor{mattgreen!80} High & 0.85& \mybar{0.0085}\\
\cellcolor{mattred!80} Low & \cellcolor{mattgreen!80} High & \cellcolor{mattred!80} Low & \cellcolor{mattgreen!80} High & \cellcolor{mattgreen!80} High & \cellcolor{mattgreen!80} High & 0.79& \mybar{0.0079}\\
\cellcolor{mattgreen!80} High & \cellcolor{mattgreen!80} High & \cellcolor{mattgreen!80} High & \cellcolor{mattred!80} Low & \cellcolor{mattred!80} Low & \cellcolor{mattred!80} Low & 0.64& \mybar{0.0064}\\
\cellcolor{mattred!80} Low & \cellcolor{mattred!80} Low & \cellcolor{mattred!80} Low & \cellcolor{mattgreen!80} High & \cellcolor{mattgreen!80} High & \cellcolor{mattgreen!80} High & 0.52& \mybar{0.0052}\\
\cellcolor{mattgreen!80} High & \cellcolor{mattgreen!80} High & \cellcolor{mattgreen!80} High & \cellcolor{mattgreen!80} High & \cellcolor{mattred!80} Low & \cellcolor{mattgreen!80} High & 0.49& \mybar{0.0049}\\
\cellcolor{mattgreen!80} High & \cellcolor{mattred!80} Low & \cellcolor{mattgreen!80} High & \cellcolor{mattred!80} Low & \cellcolor{mattred!80} Low & \cellcolor{mattred!80} Low & 0.35& \mybar{0.0035}\\
\cellcolor{mattgreen!80} High & \cellcolor{mattred!80} Low & \cellcolor{mattgreen!80} High & \cellcolor{mattred!80} Low & \cellcolor{mattgreen!80} High & \cellcolor{mattred!80} Low & 0.30& \mybar{0.003}\\
\cellcolor{mattgreen!80} High & \cellcolor{mattred!80} Low & \cellcolor{mattgreen!80} High & \cellcolor{mattgreen!80} High & \cellcolor{mattgreen!80} High & \cellcolor{mattred!80} Low & 0.26& \mybar{0.0026}\\
\cellcolor{mattgreen!80} High & \cellcolor{mattred!80} Low & \cellcolor{mattred!80} Low & \cellcolor{mattgreen!80} High & \cellcolor{mattred!80} Low & \cellcolor{mattgreen!80} High & 0.25& \mybar{0.0025}\\
\cellcolor{mattgreen!80} High & \cellcolor{mattgreen!80} High & \cellcolor{mattgreen!80} High & \cellcolor{mattgreen!80} High & \cellcolor{mattgreen!80} High & \cellcolor{mattred!80} Low & 0.23& \mybar{0.0023}\\
\cellcolor{mattred!80} Low & \cellcolor{mattred!80} Low & \cellcolor{mattred!80} Low & \cellcolor{mattred!80} Low & \cellcolor{mattred!80} Low & \cellcolor{mattgreen!80} High & 0.23& \mybar{0.0023}\\
\cellcolor{mattred!80} Low & \cellcolor{mattred!80} Low & \cellcolor{mattred!80} Low & \cellcolor{mattred!80} Low & \cellcolor{mattgreen!80} High & \cellcolor{mattred!80} Low & 0.21& \mybar{0.0021}\\
\cellcolor{mattgreen!80} High & \cellcolor{mattgreen!80} High & \cellcolor{mattgreen!80} High & \cellcolor{mattred!80} Low & \cellcolor{mattgreen!80} High & \cellcolor{mattred!80} Low & 0.21& \mybar{0.0021}\\
\cellcolor{mattred!80} Low & \cellcolor{mattred!80} Low & \cellcolor{mattred!80} Low & \cellcolor{mattgreen!80} High & \cellcolor{mattred!80} Low & \cellcolor{mattgreen!80} High & 0.20& \mybar{0.002}\\
\cellcolor{mattred!80} Low & \cellcolor{mattgreen!80} High & \cellcolor{mattred!80} Low & \cellcolor{mattred!80} Low & \cellcolor{mattred!80} Low & \cellcolor{mattgreen!80} High & 0.13& \mybar{0.0013}\\
\cellcolor{mattred!80} Low & \cellcolor{mattgreen!80} High & \cellcolor{mattgreen!80} High & \cellcolor{mattred!80} Low & \cellcolor{mattgreen!80} High & \cellcolor{mattred!80} Low & 0.11& \mybar{0.0011}\\
\cellcolor{mattred!80} Low & \cellcolor{mattgreen!80} High & \cellcolor{mattred!80} Low & \cellcolor{mattgreen!80} High & \cellcolor{mattred!80} Low & \cellcolor{mattgreen!80} High & 0.10& \mybar{0.001}\\
\cellcolor{mattred!80} Low & \cellcolor{mattgreen!80} High & \cellcolor{mattred!80} Low & \cellcolor{mattred!80} Low & \cellcolor{mattgreen!80} High & \cellcolor{mattred!80} Low & 0.09& \mybar{0.0009}\\
\cellcolor{mattred!80} Low & \cellcolor{mattgreen!80} High & \cellcolor{mattred!80} Low & \cellcolor{mattred!80} Low & \cellcolor{mattgreen!80} High & \cellcolor{mattgreen!80} High & 0.09& \mybar{0.0009}\\
\cellcolor{mattgreen!80} High & \cellcolor{mattred!80} Low & \cellcolor{mattred!80} Low & \cellcolor{mattgreen!80} High & \cellcolor{mattred!80} Low & \cellcolor{mattred!80} Low & 0.09& \mybar{0.0009}\\
\cellcolor{mattgreen!80} High & \cellcolor{mattgreen!80} High & \cellcolor{mattgreen!80} High & \cellcolor{mattgreen!80} High & \cellcolor{mattred!80} Low & \cellcolor{mattred!80} Low & 0.08& \mybar{0.0008}\\
\cellcolor{mattred!80} Low & \cellcolor{mattgreen!80} High & \cellcolor{mattgreen!80} High & \cellcolor{mattred!80} Low & \cellcolor{mattgreen!80} High & \cellcolor{mattgreen!80} High & 0.08& \mybar{0.0008}\\
\cellcolor{mattgreen!80} High & \cellcolor{mattred!80} Low & \cellcolor{mattred!80} Low & \cellcolor{mattred!80} Low & \cellcolor{mattred!80} Low & \cellcolor{mattgreen!80} High & 0.07& \mybar{0.0007}\\
\cellcolor{mattgreen!80} High & \cellcolor{mattgreen!80} High & \cellcolor{mattred!80} Low & \cellcolor{mattred!80} Low & \cellcolor{mattred!80} Low & \cellcolor{mattred!80} Low & 0.07& \mybar{0.0007}\\
\cellcolor{mattred!80} Low & \cellcolor{mattred!80} Low & \cellcolor{mattred!80} Low & \cellcolor{mattred!80} Low & \cellcolor{mattgreen!80} High & \cellcolor{mattgreen!80} High & 0.07& \mybar{0.0007}\\
\cellcolor{mattgreen!80} High & \cellcolor{mattgreen!80} High & \cellcolor{mattgreen!80} High & \cellcolor{mattred!80} Low & \cellcolor{mattred!80} Low & \cellcolor{mattgreen!80} High & 0.06& \mybar{0.0006}\\
\end{tabular} }
\end{center}
\end{table}

\begin{table}[H]
\caption{\label{tab:dec2008}Proportions of callers in categories defined by the six new measures of mobility from Groups A, B and C in December 2008. The measures are evaluated based on the spatiotemporal trajectories of each of the 1,034,431 persons that used a major cellular phone services provider's Rwandan network  during that month. From the $2^6=64$ possible categories, only the $46$ categories that contain at least $0.05\%$ of the callers from that month are shown.}
\begin{center} {\small\renewcommand{\arraystretch}{0.5}
\begin{tabular}{c|cc|ccc|rl}
{\bf Group A}& \multicolumn{2}{c|}{\bf Group B} & \multicolumn{3}{c|}{\bf Group C} & \multicolumn{2}{c}{\bf Percentage}\\ 
 NT &  GCV-R & SV-R & DT-R & TT-R & GCV-R & \\ \hline
\cellcolor{mattred!80} Low & \cellcolor{mattred!80} Low & \cellcolor{mattred!80} Low & \cellcolor{mattred!80} Low & \cellcolor{mattred!80} Low & \cellcolor{mattred!80} Low & 31.66& \mybar{0.3166}\\
\cellcolor{mattgreen!80} High & \cellcolor{mattgreen!80} High & \cellcolor{mattgreen!80} High & \cellcolor{mattgreen!80} High & \cellcolor{mattgreen!80} High & \cellcolor{mattgreen!80} High & 28.69& \mybar{0.2869}\\
\cellcolor{mattgreen!80} High & \cellcolor{mattred!80} Low & \cellcolor{mattred!80} Low & \cellcolor{mattgreen!80} High & \cellcolor{mattgreen!80} High & \cellcolor{mattgreen!80} High & 7.28& \mybar{0.0728}\\
\cellcolor{mattred!80} Low & \cellcolor{mattgreen!80} High & \cellcolor{mattgreen!80} High & \cellcolor{mattred!80} Low & \cellcolor{mattred!80} Low & \cellcolor{mattred!80} Low & 7.06& \mybar{0.0706}\\
\cellcolor{mattgreen!80} High & \cellcolor{mattred!80} Low & \cellcolor{mattgreen!80} High & \cellcolor{mattgreen!80} High & \cellcolor{mattgreen!80} High & \cellcolor{mattgreen!80} High & 3.00& \mybar{0.03}\\
\cellcolor{mattred!80} Low & \cellcolor{mattgreen!80} High & \cellcolor{mattgreen!80} High & \cellcolor{mattgreen!80} High & \cellcolor{mattgreen!80} High & \cellcolor{mattgreen!80} High & 3.00& \mybar{0.03}\\
\cellcolor{mattgreen!80} High & \cellcolor{mattred!80} Low & \cellcolor{mattred!80} Low & \cellcolor{mattred!80} Low & \cellcolor{mattred!80} Low & \cellcolor{mattred!80} Low & 2.90& \mybar{0.029}\\
\cellcolor{mattgreen!80} High & \cellcolor{mattgreen!80} High & \cellcolor{mattred!80} Low & \cellcolor{mattgreen!80} High & \cellcolor{mattgreen!80} High & \cellcolor{mattgreen!80} High & 2.27& \mybar{0.0227}\\
\cellcolor{mattred!80} Low & \cellcolor{mattgreen!80} High & \cellcolor{mattred!80} Low & \cellcolor{mattred!80} Low & \cellcolor{mattred!80} Low & \cellcolor{mattred!80} Low & 2.01& \mybar{0.0201}\\
\cellcolor{mattred!80} Low & \cellcolor{mattred!80} Low & \cellcolor{mattgreen!80} High & \cellcolor{mattred!80} Low & \cellcolor{mattred!80} Low & \cellcolor{mattred!80} Low & 1.63& \mybar{0.0163}\\
\cellcolor{mattred!80} Low & \cellcolor{mattgreen!80} High & \cellcolor{mattgreen!80} High & \cellcolor{mattgreen!80} High & \cellcolor{mattred!80} Low & \cellcolor{mattgreen!80} High & 1.02& \mybar{0.0102}\\
\cellcolor{mattgreen!80} High & \cellcolor{mattred!80} Low & \cellcolor{mattred!80} Low & \cellcolor{mattgreen!80} High & \cellcolor{mattgreen!80} High & \cellcolor{mattred!80} Low & 0.88& \mybar{0.0088}\\
\cellcolor{mattred!80} Low & \cellcolor{mattgreen!80} High & \cellcolor{mattred!80} Low & \cellcolor{mattgreen!80} High & \cellcolor{mattgreen!80} High & \cellcolor{mattgreen!80} High & 0.88& \mybar{0.0088}\\
\cellcolor{mattgreen!80} High & \cellcolor{mattred!80} Low & \cellcolor{mattred!80} Low & \cellcolor{mattred!80} Low & \cellcolor{mattgreen!80} High & \cellcolor{mattred!80} Low & 0.87& \mybar{0.0087}\\
\cellcolor{mattred!80} Low & \cellcolor{mattred!80} Low & \cellcolor{mattred!80} Low & \cellcolor{mattgreen!80} High & \cellcolor{mattgreen!80} High & \cellcolor{mattgreen!80} High & 0.84& \mybar{0.0084}\\
\cellcolor{mattgreen!80} High & \cellcolor{mattgreen!80} High & \cellcolor{mattgreen!80} High & \cellcolor{mattred!80} Low & \cellcolor{mattred!80} Low & \cellcolor{mattred!80} Low & 0.71& \mybar{0.0071}\\
\cellcolor{mattred!80} Low & \cellcolor{mattgreen!80} High & \cellcolor{mattgreen!80} High & \cellcolor{mattred!80} Low & \cellcolor{mattred!80} Low & \cellcolor{mattgreen!80} High & 0.64& \mybar{0.0064}\\
\cellcolor{mattgreen!80} High & \cellcolor{mattgreen!80} High & \cellcolor{mattgreen!80} High & \cellcolor{mattgreen!80} High & \cellcolor{mattred!80} Low & \cellcolor{mattgreen!80} High & 0.46& \mybar{0.0046}\\
\cellcolor{mattgreen!80} High & \cellcolor{mattred!80} Low & \cellcolor{mattgreen!80} High & \cellcolor{mattred!80} Low & \cellcolor{mattred!80} Low & \cellcolor{mattred!80} Low & 0.39& \mybar{0.0039}\\
\cellcolor{mattred!80} Low & \cellcolor{mattred!80} Low & \cellcolor{mattred!80} Low & \cellcolor{mattred!80} Low & \cellcolor{mattgreen!80} High & \cellcolor{mattred!80} Low & 0.35& \mybar{0.0035}\\
\cellcolor{mattgreen!80} High & \cellcolor{mattgreen!80} High & \cellcolor{mattgreen!80} High & \cellcolor{mattgreen!80} High & \cellcolor{mattgreen!80} High & \cellcolor{mattred!80} Low & 0.26& \mybar{0.0026}\\
\cellcolor{mattgreen!80} High & \cellcolor{mattred!80} Low & \cellcolor{mattgreen!80} High & \cellcolor{mattgreen!80} High & \cellcolor{mattgreen!80} High & \cellcolor{mattred!80} Low & 0.25& \mybar{0.0025}\\
\cellcolor{mattgreen!80} High & \cellcolor{mattred!80} Low & \cellcolor{mattred!80} Low & \cellcolor{mattgreen!80} High & \cellcolor{mattred!80} Low & \cellcolor{mattgreen!80} High & 0.24& \mybar{0.0024}\\
\cellcolor{mattgreen!80} High & \cellcolor{mattred!80} Low & \cellcolor{mattgreen!80} High & \cellcolor{mattred!80} Low & \cellcolor{mattgreen!80} High & \cellcolor{mattred!80} Low & 0.21& \mybar{0.0021}\\
\cellcolor{mattred!80} Low & \cellcolor{mattred!80} Low & \cellcolor{mattred!80} Low & \cellcolor{mattgreen!80} High & \cellcolor{mattred!80} Low & \cellcolor{mattgreen!80} High & 0.20& \mybar{0.002}\\
\cellcolor{mattgreen!80} High & \cellcolor{mattgreen!80} High & \cellcolor{mattgreen!80} High & \cellcolor{mattred!80} Low & \cellcolor{mattgreen!80} High & \cellcolor{mattred!80} Low & 0.16& \mybar{0.0016}\\
\cellcolor{mattred!80} Low & \cellcolor{mattred!80} Low & \cellcolor{mattred!80} Low & \cellcolor{mattred!80} Low & \cellcolor{mattred!80} Low & \cellcolor{mattgreen!80} High & 0.16& \mybar{0.0016}\\
\cellcolor{mattgreen!80} High & \cellcolor{mattred!80} Low & \cellcolor{mattred!80} Low & \cellcolor{mattred!80} Low & \cellcolor{mattgreen!80} High & \cellcolor{mattgreen!80} High & 0.14& \mybar{0.0014}\\
\cellcolor{mattred!80} Low & \cellcolor{mattgreen!80} High & \cellcolor{mattred!80} Low & \cellcolor{mattred!80} Low & \cellcolor{mattgreen!80} High & \cellcolor{mattred!80} Low & 0.14& \mybar{0.0014}\\
\cellcolor{mattred!80} Low & \cellcolor{mattgreen!80} High & \cellcolor{mattgreen!80} High & \cellcolor{mattred!80} Low & \cellcolor{mattgreen!80} High & \cellcolor{mattred!80} Low & 0.13& \mybar{0.0013}\\
\cellcolor{mattred!80} Low & \cellcolor{mattgreen!80} High & \cellcolor{mattred!80} Low & \cellcolor{mattgreen!80} High & \cellcolor{mattred!80} Low & \cellcolor{mattgreen!80} High & 0.12& \mybar{0.0012}\\
\cellcolor{mattred!80} Low & \cellcolor{mattgreen!80} High & \cellcolor{mattgreen!80} High & \cellcolor{mattred!80} Low & \cellcolor{mattgreen!80} High & \cellcolor{mattgreen!80} High & 0.12& \mybar{0.0012}\\
\cellcolor{mattgreen!80} High & \cellcolor{mattred!80} Low & \cellcolor{mattred!80} Low & \cellcolor{mattred!80} Low & \cellcolor{mattred!80} Low & \cellcolor{mattgreen!80} High & 0.11& \mybar{0.0011}\\
\cellcolor{mattred!80} Low & \cellcolor{mattgreen!80} High & \cellcolor{mattred!80} Low & \cellcolor{mattred!80} Low & \cellcolor{mattred!80} Low & \cellcolor{mattgreen!80} High & 0.11& \mybar{0.0011}\\
\cellcolor{mattred!80} Low & \cellcolor{mattred!80} Low & \cellcolor{mattred!80} Low & \cellcolor{mattgreen!80} High & \cellcolor{mattgreen!80} High & \cellcolor{mattred!80} Low & 0.10& \mybar{0.001}\\
\cellcolor{mattgreen!80} High & \cellcolor{mattred!80} Low & \cellcolor{mattred!80} Low & \cellcolor{mattgreen!80} High & \cellcolor{mattred!80} Low & \cellcolor{mattred!80} Low & 0.10& \mybar{0.001}\\
\cellcolor{mattred!80} Low & \cellcolor{mattgreen!80} High & \cellcolor{mattred!80} Low & \cellcolor{mattred!80} Low & \cellcolor{mattgreen!80} High & \cellcolor{mattgreen!80} High & 0.09& \mybar{0.0009}\\
\cellcolor{mattred!80} Low & \cellcolor{mattred!80} Low & \cellcolor{mattred!80} Low & \cellcolor{mattred!80} Low & \cellcolor{mattgreen!80} High & \cellcolor{mattgreen!80} High & 0.09& \mybar{0.0009}\\
\cellcolor{mattgreen!80} High & \cellcolor{mattgreen!80} High & \cellcolor{mattgreen!80} High & \cellcolor{mattred!80} Low & \cellcolor{mattred!80} Low & \cellcolor{mattgreen!80} High & 0.08& \mybar{0.0008}\\
\cellcolor{mattgreen!80} High & \cellcolor{mattgreen!80} High & \cellcolor{mattgreen!80} High & \cellcolor{mattgreen!80} High & \cellcolor{mattred!80} Low & \cellcolor{mattred!80} Low & 0.08& \mybar{0.0008}\\
\cellcolor{mattred!80} Low & \cellcolor{mattred!80} Low & \cellcolor{mattgreen!80} High & \cellcolor{mattgreen!80} High & \cellcolor{mattred!80} Low & \cellcolor{mattgreen!80} High & 0.05& \mybar{0.0005}\\
\cellcolor{mattgreen!80} High & \cellcolor{mattgreen!80} High & \cellcolor{mattred!80} Low & \cellcolor{mattgreen!80} High & \cellcolor{mattred!80} Low & \cellcolor{mattgreen!80} High & 0.05& \mybar{0.0005}\\
\cellcolor{mattgreen!80} High & \cellcolor{mattgreen!80} High & \cellcolor{mattred!80} Low & \cellcolor{mattred!80} Low & \cellcolor{mattred!80} Low & \cellcolor{mattred!80} Low & 0.05& \mybar{0.0005}\\
\cellcolor{mattred!80} Low & \cellcolor{mattgreen!80} High & \cellcolor{mattgreen!80} High & \cellcolor{mattgreen!80} High & \cellcolor{mattgreen!80} High & \cellcolor{mattred!80} Low & 0.05& \mybar{0.0005}\\
\cellcolor{mattred!80} Low & \cellcolor{mattred!80} Low & \cellcolor{mattgreen!80} High & \cellcolor{mattgreen!80} High & \cellcolor{mattgreen!80} High & \cellcolor{mattgreen!80} High & 0.05& \mybar{0.0005}\\
\cellcolor{mattgreen!80} High & \cellcolor{mattgreen!80} High & \cellcolor{mattgreen!80} High & \cellcolor{mattred!80} Low & \cellcolor{mattgreen!80} High & \cellcolor{mattgreen!80} High & 0.05& \mybar{0.0005}\\
\end{tabular} }
\end{center}
\end{table}

\subsection*{SI7: Checking the Quality of CDRs with the Measure of Mobility Time Traveled (TT-R)}

The measure of mobility time traveled (TT-R) can be used to filter out spatiotemporal trajectories that might have been adversely affected by errors in the cellular service provider's databases of CDRs. Such trajectories can also arise if intruders gain unauthorized access to mobile phones and use them to communicate at the same time as the actual owners. CDRs generated by an intruder and an owner of a mobile phone are saved in the same spatiotemporal trajectory. Measures of mobility for trajectories generated by two or more users of the same phone who aret not located close to each other yield unusually high mobility levels. The TT-R measure can be especially useful in identifying such unusual trajectories that might need to be discarded from population mobility studies.

Here we use the values of TT-R to identify the spatiotemporal trajectories with an average travel time of more than 12 hours per day. We refer to those trajectories with large values of TT-R as unusual, otherwise a trajectory is considered to be normal. In Table \ref{tab:qualitycheck}  we report the number of normal and the number of unusual trajectories  for each of the 44 months of Rwandan CDRs. Despite monthly fluctuations, the percentage of unusual trajectories in a given month never exceed $0.5\%$ of the total number of callers for that month, with only one spike in December 2006. The first and third quartiles, the means and the medians of TT-R appear consistent over time and are not large (~500 minutes or ~16 travel hours/day). The largest maximum for any month period is large (5000 or ~166 travel hours/day), but these extreme outliers represent a small proportion of all the unusual monthly spatiotemporal trajectories, and a very small proportion of the overall dataset. 

\begin{table}[H]
\caption{\label{tab:qualitycheck}Identification of the monthly spatiotemporal trajectories with an average travel time of more than 12 hours per day based on the values of the measure of mobility time traveled (TT-R).}
\begin{center} {\small
\setlength{\tabcolsep}{2pt}
\renewcommand{\arraystretch}{0.85}
\begin{tabular}{cccccccccc} 
Month/ & No. of  Normal  & No. of  Unusual & $\%$ &  \multicolumn{6}{c}{Summary}\\ \cline{5-10}
Year & Trajectories & Trajectories &  & Min. & 1st Qu. & Median & Mean & 3rd Qu. &  Max.\\ \hline
Jun 2005  &  190634  &  6  &  $<0.01$  &  362.5 & 382.1 & 422.3 & 440 & 439.5 & 618.8\\ 
Jul 2005  &  200160  &  5  &  $<0.01$ &  392.4 & 416.8 & 740.4 & 656.1 & 848.1 & 883\\
Aug 2005  &  208136  &  4  &  $<0.01$  &  380 & 572.9 & 703.2 & 700.4 & 830.8 & 1015\\
Sep 2005  &  211767  &  5  &  $<0.01$  &  404.6 & 406.5 & 643.8 & 614.1 & 765 & 850.9\\
Oct 2005  &  209705  &  3  &  $<0.01$  &  376.7 & 442 & 507.3 & 506.7 & 571.6 & 635.9\\
Nov 2005  &  226194  &  113  &  0.05  &  360.8 & 414.3 & 490.1 & 541.3 & 606.1 & 1217\\ 
Dec 2005  &  238572  &  138  &  0.06  &  374.7 & 448.9 & 521.2 & 600.8 & 674.1 & 1755\\
Jan 2006  &  244138  &  123  &  0.05  &  373 & 431 & 488.8 & 570.7 & 637.5 & 1555\\
Feb 2006  &  243243  &  99  &  0.04  &  337.4 & 375.8 & 426 & 497.6 & 562.1 & 1266\\
Mar 2006  &  249298  &  79  &  0.03  &  372.4 & 418.9 & 486.2 & 549.6 & 610.8 & 1508\\ 
Apr 2006  &  253119  &  123  &  0.05  &  364.5 & 400.2 & 461.2 & 531.7 & 573.7 & 1675\\
May 2006  &  254997  &  242  &  0.09  &  372.1 & 407.7 & 465.3 & 524.2 & 582.5 & 1737\\ 
Jun 2006  &  256424  &  390  &  0.15  &  360.5 & 400.2 & 461.1 & 522.2 & 575.8 & 3373\\
Jul 2006  &  262474  &  440  &  0.17  &  372.1 & 408.1 & 462.4 & 513.5 & 557.8 & 3734\\
Aug 2006  &  270738  &  782  &  0.29  &  372.1 & 406.2 & 465.3 & 524.5 & 573.2 & 4298\\
Sep 2006  &  272543  &  925  &  0.34  &  360.2 & 394 & 450.9 & 507.7 & 558.7 & 3417\\
Oct 2006  &  280419  &  972  &  0.35  &  372.1 & 413.4 & 471.7 & 533.4 & 585.5 & 3658\\
Nov 2006  &  286481  &  944  &  0.33  &  360 & 399.5 & 457.6 & 509.1 & 552.3 & 3490\\
Dec 2006  &  310877  &  1316  &  0.42  &  372 & 409.6 & 471.7 & 528 & 576.5 & 2510\\
Jan 2007  &  320249  &  1262  &  0.39  &  372 & 409.8 & 465.8 & 527.1 & 570.5 & 2565\\
Feb 2007  &  322645  &  1083  &  0.33  &  336.1 & 374.3 & 427.2 & 486.9 & 526.4 & 2838\\ 
Mar 2007  &  338276  &  1206  &  0.36  &  372 & 414.4 & 472.9 & 536.8 & 580.5 & 2610\\
Apr 2007  &  354874  &  1112  &  0.31  &  360 & 398.6 & 453.2 & 515.7 & 555.4 & 2307\\
May 2007  &  375186  &  1071  &  0.28  &  372.1 & 414.3 & 466.3 & 532.5 & 576.2 & 2563\\ 
Jun 2007  &  406896  &  915  &  0.22  &  360.2 & 398.7 & 459 & 526.2 & 571.9 & 2386\\
Jul 2007  &  432598  &  831  &  0.19  &  372 & 409.6 & 467.4 & 538.4 & 581.3 & 3398\\
Aug 2007  &  456355  &  937  &  0.2  &  372.1 & 413 & 466.7 & 543.7 & 581.5 & 4037\\
Sep 2007  &  472683  &  746  &  0.16  &  360.3 & 403.1 & 460.1 & 530.7 & 569.9 & 3566\\ 
Oct 2007  &  500802  &  1033  &  0.21  &  372.2 & 411 & 469.9 & 536.5 & 584.9 & 2204\\
Nov 2007  &  523857  &  1250  &  0.24  &  360 & 394.5 & 452.7 & 517.9 & 558.1 & 2671\\
Dec 2007  &  552041  &  1486  &  0.27  &  372 & 413.8 & 473.3 & 541.9 & 587.7 & 3235\\
Jan 2008  &  583890  &  1500  &  0.26  &  372.1 & 414.5 & 481.3 & 549 & 587.1 & 2362\\
Feb 2008  &  617341  &  1528  &  0.25  &  348.2 & 386.5 & 447 & 509.7 & 553.8 & 2364\\
Mar 2008  &  653192  &  603  &  0.09  &  372 & 399.7 & 438.6 & 490 & 518.3 & 2279\\
Apr 2008  &  675318  &  657  &  0.1  &  360.2 & 388.7 & 431.3 & 497.6 & 527.4 & 2331\\ 
May 2008  &  703200  &  773  &  0.11  &  372.2 & 400.8 & 445.7 & 520.1 & 543.8 & 2952\\
Jun 2008  &  736491  &  948  &  0.13  &  360 & 392.8 & 434.3 & 506.2 & 529.4 & 2884\\ 
Jul 2008  &  787210  &  1050  &  0.13  &  372.1 & 404 & 451.5 & 511.6 & 539.4 & 2783\\
Aug 2008  &  831993  &  1400  &  0.17  &  372 & 403.3 & 448.9 & 522.1 & 542.9 & 4851\\
Sep 2008  &  864810  &  1358  &  0.16  &  360.1 & 389.7 & 434.9 & 500.4 & 525.1 & 4633\\
Oct 2008  &  903273  &  1361  &  0.15  &  372 & 402.1 & 453.1 & 517.1 & 534.6 & 4479\\
Nov 2008  &  941894  &  1249  &  0.13  &  360.1 & 387.9 & 431.2 & 504.2 & 516.3 & 5285\\ 
Dec 2008  &  1034431  &  1728  &  0.17  &  372.2 & 403.8 & 456.8 & 530.3 & 555.9 & 4186\\
Jan 2009  &  1080547  &  2122  &  0.2  &  372 & 413.4 & 485.6 & 561.8 & 611.2 & 4147\\ \hline
\end{tabular} }
\end{center}
\end{table}

\end{document}